\documentclass[twocolumn,showpacs,aps,epsfig]{revtex4}

\usepackage{graphicx}
\usepackage{epstopdf}
\usepackage{latexsym}

\usepackage[center]{subfigure}

\begin{document}

 \newcommand{\bq}{\begin{equation}}
 \newcommand{\eq}{\end{equation}}
 \newcommand{\bqn}{\begin{eqnarray}}
 \newcommand{\eqn}{\end{eqnarray}}
 \newcommand{\nb}{\nonumber}
 \newcommand{\lb}{\label}
\newcommand{\PRL}{Phys. Rev. Lett.}
\newcommand{\PL}{Phys. Lett.}
\newcommand{\PR}{Phys. Rev.}
\newcommand{\CQG}{Class. Quantum Grav.}
 
\title{Colliding branes and formation of spacetime singularities in string theory}
\author{Andreas Tziolas}
\email{Andreas_Tziolas@baylor.edu}
\affiliation{GCAP-CASPER, Physics Department, Baylor University, Waco, TX 76798, USA}
\author{Anzhong Wang}
\email{Anzhong_Wang@baylor.edu}
\affiliation{GCAP-CASPER, Physics Department, Baylor University, Waco, TX 76798, USA} 
\author{Zhong Chao Wu}
\email{zcwu@zjut.edu.cn}
\affiliation{Department of Physics, Zhejiang University of Technology, Hangzhou, 
310032, China}
\date{\today}
 
\begin{abstract}

 Colliding branes without $Z_{2}$ symmetry and the formation of spacetime singularities
 in string theory are studied. After developing the general formulas to describe such
 events, we study a particular class of exact solutions first in the 5-dimensional
 effective theory, and then lift it to the 10-dimensional spacetime. In general,
 the 5-dimensional spacetime is singular, due to the mutual focus of the two 
 colliding 3-branes. Non-singular cases also exist, but with the price that both of the
 colliding branes violate all the three energy conditions, weak, dominant, and strong.
 After lifted to 10 dimensions,  we find that the spacetime remains singular,
 whenever it is singular in the 5-dimensional effective theory. In the cases where
 no singularities are formed after the collision, we find that the two 8-branes necessarily
 violate all the energy conditions.

\end{abstract}

 
\maketitle

\section{Introduction}

Branes in string/M-Theory are fundamental constituents \cite{strings}, and of 
particular relevance to cosmology \cite{JST02, branes}. These substances can move
freely in the bulk, collide, recoil, reconnect, and whereby, among other possibilities,
form a brane gas in the early universe \cite{branegas}, or  create an ekpyrotic/cyclic universe 
\cite{cyc}. Understanding these processes is fundamental to both string/M-Theory
and their applications to cosmology \cite{Wu83}.

Recently, Maeda and his collaborators numerically studied  the collision 
of two branes in a five-dimensional bulk, and found that the formation of a spacelike
singularity after the collision is generic \cite{MT04} (See also \cite{BP03}). 
This is an important result, as it implies that a low-energy description of colliding branes 
breaks down at some point, and a complete predictability is lost, without the complete
theory of quantum gravity. Similar conclusions were obtained from the studies of two 
colliding orbifold branes \cite{chen06}. However, lately it was argued that, from the 
point of view of the higher dimensional spacetime where the low effective action was
derived, these singularities are very mild and can be easily regularised \cite{LFT06}.

Lately, we constructed a class of exact solutions  with two free parameters 
to the five-dimensional Einstein field equations, which represents
the collision of two timelike 3-branes \cite{TW08}. We found that, among other things,  
spacelike  singularities generically develop after the collision, due to the 
mutual focus of the two branes. Non-singular spacetime can be constructed only 
in the case where both of the two branes violate the energy conditions.

In this paper, we shall systematically study the collision of two timelike 8-branes
without $Z_{2}$ symmetry in the framework of string theory. In particular, in Section
II, starting with the  Neveu-Schwarz/Neveu-Schwarz (NS-NS) sector  in (D+d) dimensions, 
$\hat{M}_{D+d} = M_{D}\times M_{d}$, we first obtain the D-dimensional effective
theory in both the string frame and the Einstein frame,
by toroidal compactifications. To study the collision of two branes, we add
brane actions to the D-dimensional effective action, and then derive the gravitational
and matter field equations, including the ones on the two branes. In Section III,
we apply these general formulas to the case where $D = 5 = d$ for a large class of 
spacetimes, and obtain the explicit field equations both outside the two branes and 
on the two branes. In Section IV, we construct a  class of exact solutions in the 
Einstein frame, in which the potential of the radion field on the two branes take an 
exponential form, while the matter fields on the two branes are dust fluids. After 
identifying spacetime singularities both outside and on the branes, we are able to 
draw the corresponding Penrose diagrams for various cases.  In Sections V, we study
the local and global properties of these solutions in the 5-dimensional string frame,
while in Section VI we first lift the solutions to  10 dimensions, and then study
the local and global properties of these 10-dimensional solutions in details. 
In Section VII, we derive our main conclusions and present some remarks. There is
also an Appendix, in which we study a class of 10-dimensional spacetimes. In particular,
we divide  the Einstein tensor explicitly into three parts, one on each side of 
a colliding brane, and the other is on the brane. It is remarkable that the part on the 
brane can be written in the form of an anisotropic fluid.

\section{The Model}

\renewcommand{\theequation}{2.\arabic{equation}}
\setcounter{equation}{0}

Let us consider the toroidal compactification of the NS-NS sector of the action 
in (D+d) dimensions, $\hat{M}_{D+d} = M_{D}\times M_{d}$, where for the string theory we have
$D+d = 10$. Then, the action takes the form
\cite{LWC00},
\bqn
\lb{2.1}
S_{D+d} &=& - \frac{1}{2\kappa^{2}_{D+d}}
\int{d^{D+d}x\sqrt{\left|\hat{g}_{D+d}\right|}  e^{-\hat{\Phi}} \left\{
{\hat{R}}_{D+d}[\hat{g}]\right.}\nb\\
& & \left. + \hat{g}^{AB}\left(\hat{\nabla}_{A}\hat{\Phi}\right)
\left(\hat{\nabla}_{B}\hat{\Phi}\right) - \frac{1}{12}{\hat{H}}^{2}\right\},
\eqn
where in this paper we consider the $(D+d)$-dimensional spacetimes described by the
metric,
\bqn
\lb{2.2}
d{\hat{s}}^{2}_{D+d} &=& \hat{g}_{AB} dx^{A}dx^{B} = 
   \gamma_{ab}\left(x^{c}\right) dx^{a}
dx^{b} \nb\\& & + \hat{\phi}^{2}\left(x^{c}\right) 
\hat{\gamma}_{ij}\left(z^{k}\right)dz^{i} dz^{j},
\eqn
with  $\gamma_{ab}\left(x^{c}\right)$ and $\hat{\phi}
\left(x^{c}\right)$ depending only on the coordinates  $x^{a}$ of 
the spacetime $M_{D}$, and $\hat{\gamma}_{ij}\left(z^{k}\right)$ only on the internal 
coordinates $z^{k}$, where $a, b, c = 0, 1, 2, ...,
D-1$; $i, j, k = D, D+1, ..., D+d - 1$; and $A, B, C = 0, 1, 2, ...,
D+ d -1$. Assuming that matter fields are all independent of $z^{k}$, 
one finds that the internal space $M_{d}$ must be Ricci flat,
\bq
\lb{2.2a}
R[\hat{\gamma}] = 0. 
\eq
For the purpose of the current work, it is sufficient to assume 
that $M_{d}$ is a $d-$dimensional torus, $T^{d} = S^{1} \times
S^{1} \times ... \times S^{1}$.  
Then, we find that 
\bqn
\lb{2.3}
\hat{R}_{D+d}\left[\hat{g}\right] &=& R_{D}\left[{\gamma}\right] 
+ \frac{d(d-1)}{\hat{\phi}^{2}} 
\gamma^{ab}\left(\nabla_{a}\hat{\phi}\right)\left(\nabla_{b}\hat{\phi}\right)\nb\\
& & - \frac{2}{\hat{\phi}^{d}}\gamma^{ab} \left(\nabla_{a}\nabla_{b}\hat{\phi}^{d}\right).
\eqn
Ignoring the dilaton $\hat{\Phi}$  and the form fields  $\hat{H}$,
\bq
\lb{2.3a}
\hat{\Phi} = 0 = \hat{H},
\eq
we find that the integration of the action (\ref{2.1})
over the internal space yields,
\bqn
\lb{2.4}
S_{D} &=& - \frac{1}{2\kappa^{2}_{D}} \int{d^{D}x\sqrt{\left|\gamma\right|}
 \hat{\phi}^{d}\left\{R_{D}\left[\gamma\right] \right.}\nb\\
& & \left. 
+ \frac{d(d-1)}{\hat{\phi}^{2}} 
\gamma^{ab}\left(\nabla_{a}\hat{\phi}\right)\left(\nabla_{b}\hat{\phi}\right)
\right\},
\eqn
where 
\bq
\lb{2.5}
\kappa_{D}^{2} \equiv \frac{\kappa_{D+d}^{2}}{V_{s}},
\eq
and $V_{s}$ is defined as
\bq
\lb{2.6}
V_{s} \equiv \int{\sqrt{\hat{\gamma}} d^{d}z}.
\eq
For a string scale compactification, we have $V_{s} = 
\left(2\pi \sqrt{\alpha'}\right)^{d}$, where $\left(2\pi  \alpha'\right)$
is the inverse string tension.

After the conformal transformation,
\bq
\lb{2.7}
g_{ab} = \hat{\phi}^{\frac{2d}{D-2}}\gamma_{ab},
\eq
the D-dimensional effective action of Eq.(\ref{2.4}) can be cast in the
minimally coupled form,
\bqn
\lb{2.8}
S_{D}^{(E)} &=& - \frac{1}{2\kappa^{2}_{D}}\int{d^{D}x\sqrt{\left|g_{D}\right|}
  \left\{ R_{D}\left[g\right] 
-  \kappa^{2}_{D} \left(\nabla\phi\right)^{2}\right\}},\nb\\ 
\eqn
where 
\bq
\lb{2.9}
\phi \equiv \pm 
  \left(\frac{(D+d-2)d}{\kappa^{2}_{D}\left(D-2\right)}\right)^{1/2} 
  \ln\left(\hat{\phi}\right).
\eq

The action of Eq.(\ref{2.4}) is usually referred to as  {\em the string frame},
and the one of Eq.(\ref{2.8}) as  {\em the Einstein frame}. It should be noted that
solutions related by this conformal transformation can have completely different 
physical and geometrical properties in the two frames. In particular, in one 
frame a solution can be singular, while in the other it can be totally free from 
any kind of singularities. A simple example is the flat FRW universe which is 
always conformally flat, $\gamma_{ab} = a^{2}(\tau)\eta_{ab}$. But the spacetime
described by $\gamma_{ab}$ usually has a big bang singularity, while the one
described by $\eta_{ab}$ is Minkowski, and does not have any kind of spacetime 
singularities.

To study the collision of two branes, we add the  following brane actions to
$S_{D}^{(E)}$ of Eq.(\ref{2.8}),
\bqn
\lb{2.10}
S^{(E, I)}_{D-1, m} &=&  \int_{M^{(I)}_{D-1}}{\sqrt{\left|g^{(I)}_{D-1}\right|}
\left(
{\cal{L}}^{(m, I)}_{D-1}(\psi) - V^{(I)}_{D-1}(\phi)\right)}\nb\\
& & \times  d^{D-1}\xi_{(I)},
\eqn
where $I = 1, 2,\; V^{(I)}_{D-1}(\phi)$  denotes the potential of the scalar 
field $\phi$ on the I-th brane, and $\xi_{(I)}^{\mu}$'s are the intrinsic coordinates of the 
I-th brane, where $\mu, \; \nu,\; \lambda = 0, 1, 2, ..., D-2$. ${\cal{L}}^{(m, I)}_{D-1}(\psi)$
is the Lagrangian density of matter fields located on the I-th brane, denoted collectively
by $\psi$. It should be noted that the above action does not include kinetic terms of the 
scalar field on the branes. This setup is quite similar to the Horava-Witten heterotic M-Theory 
on $S^{1}/Z_{2}$ \cite{HW96,GWW07}, in which the two potentials $V^{(1)}_{4}(\phi)$
and $V^{(2)}_{4}(\phi)$ have opposite signs. It is also similar to  the modulus stabilization 
mechanism  of Goldberger and Wise \cite{GW99}, which has been lately applied to orbifold
branes in string theory \cite{WS07}. The two branes are localized 
on the surfaces,
\bq
\lb{2.11}
\Phi_{I}\left(x^{a}\right)  = 0,
\eq 
or equivalently
\bq
\lb{2.12}
x^{a} = x^{a}\left(\xi^{\mu}_{(I)}\right).
\eq 
$g^{(I)}_{D-1}$ denotes the determinant of the reduced metric  $g_{\mu\nu}^{(I)}$  of the I-th  brane, 
defined as
\bq
\lb{2.13}
g_{\mu\nu}^{(I)} \equiv \left. g_{ab} e^{(I)a}_{(\mu)} e^{(I)b}_{(\nu)}\right|_{M^{(I)}_{D-1}}, 
\eq
where
\bq
\lb{2.14}
e^{(I)\; a}_{(\mu)} \equiv \left.\frac{\partial x^{a}}{\partial \xi^{\mu}_{(I)}}\right|_{M^{(I)}_{D-1}}. 
\eq
Then,  the total action is given by,
\bq
\lb{2.15}
S^{(E)}_{total} = S_{D}^{(E)}  + \sum_{I=1}^{2}{S^{(E, I)}_{D-1, m}}.
\eq

Variation of the total action (\ref{2.15}) with respect to $g_{ab}$ yields the D-dimensional 
gravitational field equations,
\bqn
\lb{2.16}
 R_{ab} - \frac{1}{2}Rg_{ab} &=& \kappa^{2}_{D} \left(T^{\phi}_{ab} +
 \sum^{2}_{I=1}{\left(T^{(m, I)}_{\mu\nu} +  g^{(I)}_{\mu\nu}V^{(I)}_{D-1}(\phi)\right)}\right.\nb\\
& &   \left.\times e^{(I, \mu)}_{a}
e^{(I, \nu)}_{b}  \sqrt{\left|\frac{g^{(I)}_{D-1}}{g_{D}}\right|}
\delta\left(\Phi_{I}\right)\right), 
\eqn
where    
\bqn
\lb{2.17}
 T^{\phi}_{ab} &=& \nabla_{a}\phi\nabla_{b}\phi  
 - \frac{1}{2}g_{ab}\left(\nabla\phi\right)^{2},\nb\\
 T^{(m, I)}_{\mu\nu}  &=&     2 \frac{\delta{\cal{L}}^{(m, I)}_{D-1}}
  {\delta{g^{(I)\; \mu\nu}}} -  g^{(I)}_{\mu\nu}{\cal{L}}^{(m, I)}_{D-1},
\eqn
and $\nabla_{a}$ $\left(\nabla^{(I)}_{\mu}\right)$ denotes  the covariant derivative with respect 
to $g_{ab}$ $\left(g^{(I)}_{\mu\nu}\right)$.

Variation of the total action with respect to $\phi$, on the other hand,
 yields the Klein-Gordon field equations,
\bq
\lb{2.18} 
 \Box\phi =  - \sum^{2}_{I=1}{\frac{\partial V^{(I)}_{D-1}(\phi)}{\partial\phi}
  \sqrt{\left|\frac{g^{(I)}_{D-1}}{g_{D}}\right|} \; \delta\left(\Phi_{I}\right)}, 
\eq
where $\Box \equiv g^{ab}\nabla_{a} \nabla_{b}$. We also have
\bq
\lb{2.19}
\nabla^{(I)}_{\nu} T^{(m, I)\; \mu\nu} = 0.
\eq
 
Since we are mainly interested in collision of branes in the string theory, in the rest
of this paper we shall set $D = 5 = d$.

\section{Colliding  timelike 3-branes in the Einstein frame}

\renewcommand{\theequation}{3.\arabic{equation}}
\setcounter{equation}{0}

We consider the 5-dimensional spacetime in the Einstein frame described by the metric, 
\bqn
\lb{3.1}
ds^{2}_{5} &=& g_{ab} dx^{a}dx^{b}\nb\\
&=&  e^{2\sigma(t,y)}\left(dt^{2} - dy^{2}\right)
- e^{2\omega(t,y)}d\Sigma^{2}_{0},
\eqn
where $d\Sigma^{2}_{0} \equiv \left(dx^{2}\right)^{2} + \left(dx^{3}\right)^{2}
+ \left(dx^{4}\right)^{2}$, and $x^0=t$, $x^1=y$. Then, the non-vanishing components of the Ricci tensor
is given by
\bqn
\lb{3.2a}
R_{tt} &=& - \left\{3\omega_{,tt} + \sigma_{,tt} + 3\omega_{,t}\left(\omega_{,t}
            - \sigma_{,t}\right) \right.\nb\\
	    & & \left. - \sigma_{,yy} - 3\omega_{,y}\sigma_{,y}\right\},\\
\lb{3.2b}
R_{ty} &=& - 3\left\{\omega_{,ty} + \omega_{,t}\omega_{,y} 
            -   \omega_{,t}\sigma_{,y} - \omega_{,y}\sigma_{,t}\right\},\\
\lb{3.2c}	    
R_{yy} &=& - \left\{3\omega_{,yy} + \sigma_{,yy} + 3\omega_{,y}\left(\omega_{,y}
            - \sigma_{,y}\right) \right.\nb\\
	    & & \left. - \sigma_{,tt} - 3\omega_{,t}\sigma_{,t}\right\},\\
\lb{3.2d}	    
R_{ij} &=& \delta_{ij} e^{2\left(\omega - \sigma\right)}\left\{\omega_{,tt} 
        + 3{\omega_{,t}}^{2}\right.\nb\\
	& & \left. - \left(\omega_{,yy} + 3{\omega_{,y}}^{2}\right)\right\},
\eqn
where now $i,\; j = 2, 3, 4$, and $\omega_{,t} \equiv \partial\omega/\partial t$, 
etc. 

\begin{figure}
\includegraphics[width=\columnwidth]{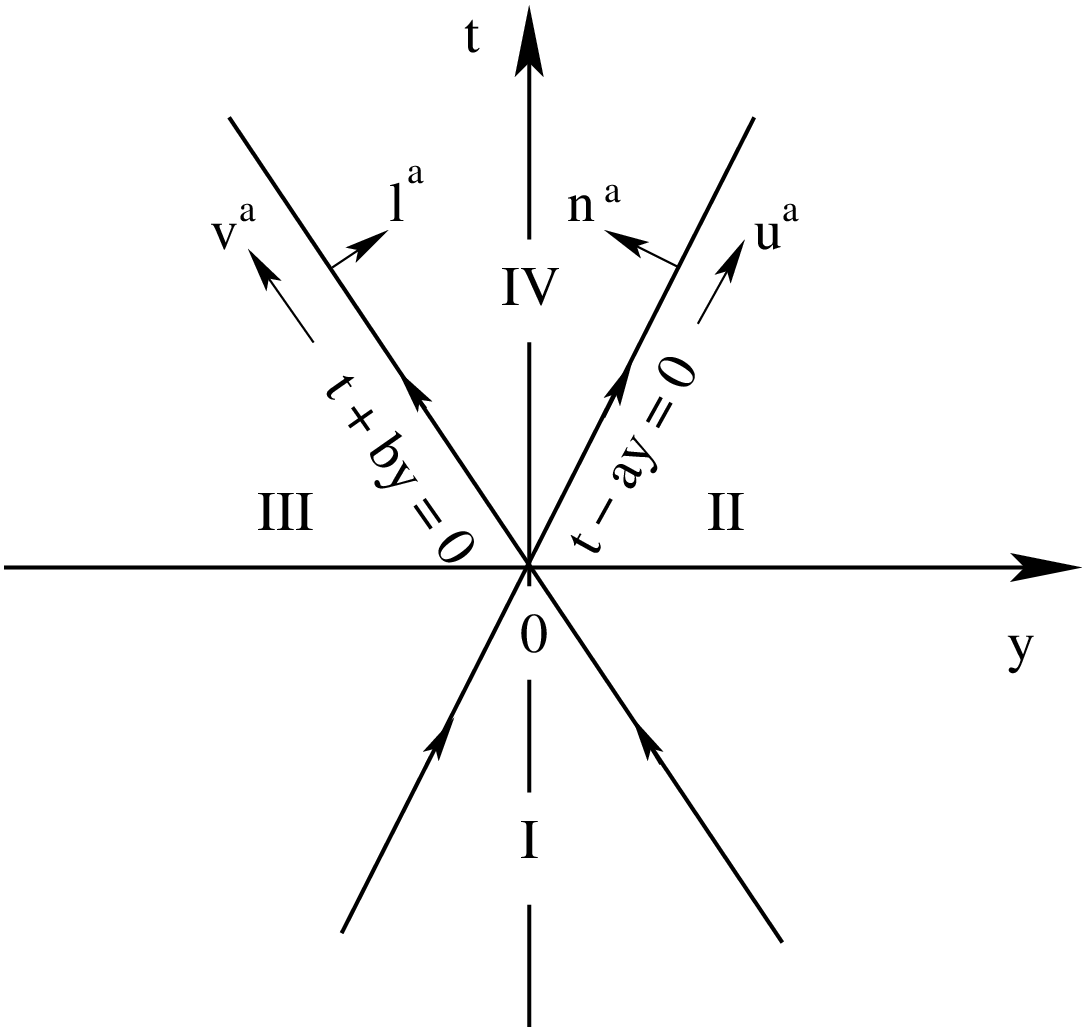}
\caption{The five-dimensional spacetime in the ($t, \; y$)-plane for 
$ a > 1,\; b  > 1$.  The two 3-branes  are moving along the hypersurfaces,
$\Sigma_{1}$ and $\Sigma_{2}$, which are defined by Eq.(\ref{3.9}) in the text. 
The four regions, $I - IV$, are defined by Eq.(\ref{3.7}).}  
\label{fig} 
\end{figure}

We assume that the two colliding 3-branes move along the hypersurfaces given,
respectively, by
\bqn
\lb{3.3}
\Phi_{1}(t, y) &=&  t - a y = 0,\nb\\
\Phi_{2}(t, y) &=&  t + b y = 0,
\eqn
where $a$ and $b$ are two arbitrary constants, subjected to the constraints,
\bq
\lb{3.4}
a^{2} > 1, \;\;\; b^{2} > 1,
\eq
in order for the two hypersurfaces to be timelike. The two colliding branes 
divide the whole spacetime into four regions, $I - IV$, which are
defined, respectively, as
\bqn
\lb{3.7}
{\mbox{Region I}} &\equiv& \left\{x^{a}: \Phi_{1} < 0, \; \Phi_{2} < 0\right\},\nb\\
{\mbox{Region II}} &\equiv& \left\{x^{a}: \Phi_{1} > 0, \; \Phi_{2} < 0\right\},\nb\\
{\mbox{Region III}} &\equiv& \left\{x^{a}: \Phi_{1} < 0, \; \Phi_{2} > 0\right\},\nb\\
{\mbox{Region IV}} &\equiv& \left\{x^{a}: \Phi_{1} > 0, \; \Phi_{2} > 0\right\},
\eqn
as shown schematically in Fig. \ref{fig}.  In each of these regions, we define
\bq
\lb{3.8}
F^{A} \equiv \left. F(t, y)\right|_{{\mbox{Region A}}},
\eq
where now $A = I, \;\ II, \; III,\; IV$. 

We also define the two hypersurfaces $\Sigma_{1}$ and $\Sigma_{2}$ as,
\bqn
\lb{3.9}
\Sigma_{1} &\equiv&  \left\{x^{a}: \; \Phi_{1} = 0\right\},\nb\\
\Sigma_{2} &\equiv&  \left\{x^{a}: \; \Phi_{2} = 0\right\}.
\eqn
Then, it can be shown that
the normal vectors to each of these two surfaces are given by
\bqn
\lb{3.5}
n_{a} &=& N \left(\delta^{t}_{a} - a \delta^{y}_{a}\right),\nb\\
l_{a} &=& L \left(\delta^{t}_{a} + b \delta^{y}_{a}\right),
\eqn
where 
\bqn
\lb{3.6}
F^{(I)} &\equiv& \left. F(t,y)\right|_{\Phi_{I} = 0},\nb\\
N &\equiv& \frac{e^{\sigma^{(1)}}}{\left(a^{2} - 1\right)^{1/2}},\nb\\
L &\equiv& \frac{e^{\sigma^{(2)}}}{\left(b^{2} - 1\right)^{1/2}},
\eqn
with $F = \left\{\sigma, \; \omega, \; \phi\right\}$. We also introduce the two timelike
vectors $u_{c}$ and $v_{c}$ via the relations,
\bqn
\lb{3.5a}
u_{a} &=& N \left(a\delta^{t}_{a} -  \delta^{y}_{a}\right),\nb\\
v_{a} &=& L \left(b\delta^{t}_{a} +  \delta^{y}_{a}\right).
\eqn
It can be shown that these vectors have the following properties,
\bqn
\lb{3.6a}
n_{a} n^{a} &=& - 1 = l_{a} l^{a},\nb\\
u_{a} u^{a} &=& + 1 = v_{a} v^{a},\nb\\
n_{a} u^{a} &=& 0 = l_{a} v^{a}.
\eqn

In the following, we shall consider field equations, (\ref{2.16}) and (\ref{2.18}),
in Regions $I - IV$ and along the hypersurfaces $\Sigma_{1,2}$, separately.

It should be noted that in the above setup, the two 3-branes do not have the $Z_{2}$
symmetry, in contrast to the setup of Horava-Witten in M theory \cite{HW96} and of
Randall-Sundrum \cite{RS1}.

\subsection{Field Equations in Regions $I - IV$}

In these regions, the field equations of Eqs.(\ref{2.16}) and (\ref{2.18}) take the
form,
\bqn
\lb{3.9a}
R^{A}_{ab} &=&  \varphi^{A}_{,a}\varphi^{A}_{,b},\\
\lb{3.9b}
\Box^{(A)}\varphi^{A} &=& 0,
\eqn
where $\varphi = \kappa_{5}\phi$, and $\Box^{(A)} \equiv g^{A\; ab}\nabla^{(A)}_{a}
\nabla^{(A)}_{b}$, and $\nabla^{(A)}_{a}$ denotes the covariant derivative with respect
to $g^{A}_{ab}$, and $g^{A}_{ab}$ is the metric defined in Region $A$. From Eq.(\ref{3.2d})
and the fact that $\varphi = \varphi(t, y)$, we find that 
\bq
\lb{3.10}
\omega = \frac{1}{3}\ln\left(f\left(t + y\right) + g\left(t- y\right)\right),
\eq
where $f\left(t + y\right)$ and  $g\left(t- y\right)$ are arbitrary functions of their 
indicated arguments. Note that in writing Eq.(\ref{3.10}) we dropped the super indices
$A$. In the following we shall adopt this convention, except for the case where
confusions may raise. In the following we consider only the case where 
\bq
\lb{3.11}
f'g' \not= 0,
\eq
where a prime denotes the ordinary derivative with respect to the indicated argument.
Then, introducing two new variables $\xi_{\pm}$ via the relations,
\bq
\lb{3.12}
\xi_{\pm}(t,y) \equiv f\left(t + y\right) \pm g\left(t- y\right),
\eq
we find that Eq.(\ref{3.9a}) yields,
\bqn
\lb{3.13a}
M_{+} &=& \frac{1}{2} \xi_{+} \left({\varphi_{+}}^{2} + {\varphi_{-}}^{2}\right),\\
\lb{3.13b}
M_{-} &=&  \xi_{+} \varphi_{+} \varphi_{-},
\eqn
and
\bq
\lb{3.14}
M_{++} - M_{--} = - \frac{1}{2}   \left({\varphi_{+}}^{2} - {\varphi_{-}}^{2}\right),
\eq
where $M_{\pm} \equiv \partial{M}/\partial{\xi_{\pm}}$, and
\bq
\lb{3.15}
M\left(\xi_{+}, \xi_{-}\right) = \sigma + \frac{1}{3}\ln\xi_{+} 
- \frac{1}{2}\ln\left(4f'g'\right).
\eq 
On the other hand, Eq.(\ref{3.9b}) can be cast in the form,
\bq
\lb{3.16}
\varphi_{++} - \varphi_{--} + \frac{1}{\xi_{+}} \varphi_{+} = 0.
\eq
It should be noted that Eqs.(\ref{3.13a})-(\ref{3.14}) and (\ref{3.16}) are not all  
independent. In fact,
Eq.(\ref{3.14}) is the integrability condition of Eqs.(\ref{3.13a}) and (\ref{3.13b}), and
can be obtained from Eqs.(\ref{3.13a}), (\ref{3.13b}) and (\ref{3.16}). Therefore, in Regions
$I-IV$, the field equations reduce to Eqs. (\ref{3.13a}), (\ref{3.13b}) and (\ref{3.16}).

To find solutions, one may first integrate Eq.(\ref{3.16}) to find $\varphi$, and then 
integrate  Eqs.(\ref{3.13a}) and (\ref{3.13b}) to find $M$. However, Eq.(\ref{3.16}) has
infinite numbers of solutions, and the corresponding general solutions of $M$ has
not been worked out yet \cite{ver93}.  Once $\varphi$ and $M$ are known, the metric coefficients 
$\sigma$ and $\omega$ are then given by
\bqn
\lb{3.17}
 \sigma &=& M  - \frac{1}{3}\ln\left(f  + g\right)
 +  \frac{1}{2}\ln\left(4f'g'\right),\nb\\
 \omega &=& \frac{1}{3}\ln\left(f + g\right).
\eqn

\subsection{Field Equations on the 3-branes}

\subsubsection{Field Equations on  the surface $\Phi_{1} = 0$}

Across the hypersurface $\Phi_{1} = 0$, for any given $C^{0}$ function $F(t, y)$, it can
be written as \cite{WCS06},
\bq
\lb{3.18}
F(t, y) = F^{+}(t, y) H\left(\Phi_{1}\right) 
+  F^{-}(t, y) \left[1 - H\left(\Phi_{1}\right)\right],
\eq
where $F^{+} \; \left(F^{-}\right)$ denotes the function $F(t, y)$ defined in the region $\Phi_{1} > 0
\; \left(\Phi_{1} < 0\right)$, and $H(x)$ denotes the heaviside function, defined as
\bq
\lb{3.19}
H(x) = \cases{1, & $x > 0$,\cr
0, & $x < 0$.\cr}
\eq
On the other hand, projecting $F_{,a}$ onto the $n_{a}$ and $u_{a}$ directions, we find
\bq
\lb{3.20}
F_{,a} =  F_{u} u_{a} - F_{n} n_{a},
\eq
where
\bq
\lb{3.21}
F_{u} \equiv u^{a}F_{,a},\;\;\;
F_{n} \equiv n^{a}F_{,a}.
\eq
Since $\left[F_{u}\right]^{-} = 0$  due to the continuity of $F$ across the branes, 
from the above expressions we find  
\bq
\lb{3.22}
\left[F_{,a}\right]^{-} = - \left[F_{n}\right]^{-} n_{a},
\eq
where
\bq
\lb{3.23}
\left[F_{,a}\right]^{-} \equiv \lim_{\Phi_{1} \rightarrow 0^{+}}
{F^{+}_{\;,a}} - \lim_{\Phi_{1} \rightarrow 0^{-}}
{F^{-}_{\;,a}}.
\eq
Then, we find that
\bqn
\lb{3.24}
F_{,t} &=& F^{+}_{\;,t}H\left(\Phi_{1}\right) 
          +  F^{-}_{\;,t} \left[1 - H\left(\Phi_{1}\right)\right],\nb\\
F_{,y} &=& F^{+}_{\;,y}H\left(\Phi_{1}\right) 
          +  F^{-}_{\;,y} \left[1 - H\left(\Phi_{1}\right)\right],\nb\\
F_{,tt} &=& F^{+}_{\;,tt}H\left(\Phi_{1}\right) 
          +  F^{-}_{\;,tt} \left[1 - H\left(\Phi_{1}\right)\right]\nb\\
	 & & - N\left[F_{n}\right]^{-} \delta\left(\Phi_{1}\right),\nb\\
F_{,ty} &=& F^{+}_{\;,ty}H\left(\Phi_{1}\right) 
          +  F^{-}_{\;,ty} \left[1 - H\left(\Phi_{1}\right)\right]\nb\\
	 & & + aN\left[F_{n}\right]^{-} \delta\left(\Phi_{1}\right),\nb\\
F_{,yy} &=& F^{+}_{\;,yy}H\left(\Phi_{1}\right) 
          +  F^{-}_{\;,yy} \left[1 - H\left(\Phi_{1}\right)\right]\nb\\
	 & & - a^{2}N\left[F_{n}\right]^{-} \delta\left(\Phi_{1}\right),
\eqn
where $\delta\left(\Phi_{1}\right)$ denotes the Dirac delta function. 
Then, we find that the Ricci tensor given by Eqs.(\ref{3.2a})-(\ref{3.2d}) can be
cast in the form,
\bqn
\lb{3.26}
R_{ab} &=& R^{+}_{\;ab} H\left(\Phi_{1}\right) 
+  R^{-}_{\; ab} \left[1 - H\left(\Phi_{1}\right)\right]\nb\\
& & + R^{Im}_{\;ab}\delta\left(\Phi_{1}\right),
\eqn 	 
where $R^{+}_{\;ab}\; \left(R^{-}_{\; ab}\right)$ is the Ricci tensor calculated in
the region $\Phi_{1} > 0\; \left(\Phi_{1} < 0\right)$, and $R^{Im}_{\;ab}$ denotes
the Ricci tensor calculated on the hypersurface $\Phi_{1} = 0$, which has the following
non-vanishing components,
\bqn
\lb{3.27}
R^{Im}_{\;tt} &=& N \left\{3\left[\omega_{n}\right]^{-} 
             - \left(a^{2} - 1\right)\left[\sigma_{n}\right]^{-}\right\},\nb\\
R^{Im}_{\;ty} &=& -3aN\left[\omega_{n}\right]^{-},\nb\\
R^{Im}_{\;yy} &=& N \left\{3a^{2}\left[\omega_{n}\right]^{-} 
             + \left(a^{2} - 1\right)\left[\sigma_{n}\right]^{-}\right\},\nb\\	     
R^{Im}_{\;ij} &=& N e^{2(\omega^{(1)} - \sigma^{(1)})} 
              \left(a^{2} - 1\right)\left[\omega_{n}\right]^{-}\delta_{ij}.	     
\eqn

On  the hypersurface $\Phi_{1} = 0$, the metric (\ref{3.1}) reduces to
\bq
\lb{3.28}
\left.ds^{2}_{5}\right|_{\Phi_{1} = 0} =  
            g^{(1)}_{\mu\nu}d\xi^{\mu}_{(1)}d\xi^{\nu}_{(1)}\nb\\
	    = d\tau^{2}  -a^{2}_{u}(\tau)d\Sigma^{2}_{0},
\eq
where $\xi^{\mu}_{(1)} \equiv \left\{\tau, \; x^{2}, \; x^{3},\; x^{4}\right\}$, and 
\bqn
\lb{3.29}
d\tau &\equiv& \epsilon_{\tau}\left(\frac{a^{2}-1}{a^{2}}\right)^{1/2} e^{\sigma^{(I)}} dt,\nb\\
a_{u}(\tau) &\equiv&   e^{\omega^{(1)}},
\eqn
with $ \epsilon_{\tau} = \pm 1$. Then, we find that
\bqn
\lb{3.30}
e^{(1)\; a}_{(\tau)} &\equiv& \frac{\partial x^{a}}{\partial\tau} 
           = \dot{t}\left(\delta^{a}_{t} + \frac{1}{a}\delta^{a}_{y}\right),\nb\\
e^{(1)\; a}_{(i)} &\equiv& \frac{\partial x^{a}}{\partial{\xi^{i}_{(1)}}} 
           =  \delta^{a}_{i},\nb\\
\sqrt{\left|\frac{g^{(1)}_{4}}{g_{5}}\right|} &=& e^{-2\sigma^{(1)}},
\eqn	   
where $i = 2,\; 3, \; 4$ and $\dot{t} \equiv dt/d\tau$. Then, the field equations
of Eq.(\ref{2.16}) can be written as
\bqn
\lb{3.31a}
\left[\omega_{n}\right]^{-} &=& \frac{\kappa^{2}_{5}e^{-\sigma^{(1)}}}
      {3\left(a^{2} -1\right)^{1/2}} \left(\rho^{(1)}_{m} + V^{(1)}_{4}\right),\;\;\;\;\;\;\;\;\\
\lb{3.31b}
2\left[\omega_{n}\right]^{-} + \left[\sigma_{n}\right]^{-} &=& 
         \frac{\kappa^{2}_{5}e^{-\sigma^{(1)}}}
      {\left(a^{2} -1\right)^{1/2}}\left(V^{(1)}_{4}-p^{(1)}_{m}\right),\;\;
\eqn
where in writing the above expressions we had assumed that $T^{(m,1)}_{\mu\nu}$ takes
the form of a perfect fluid, 
\bqn
\lb{3.32}
T^{(m,1)}_{\mu\nu} &\equiv& \left(\rho^{(1)}_{m} + p^{(1)}_{m}\right)w^{(1)}_{\mu}w^{(1)}_{\nu} 
             - p^{(1)}_{m} g^{(1)}_{\mu\nu},\nb\\
w^{(1)}_{\mu} &=& \delta^{\tau}_{\mu}.
\eqn 
Similarly, it can be shown that the Klein-Gordon equation  (\ref{2.18}) and the conservation law of
the matter fields (\ref{2.19}) on $\Sigma_{1}$ take, respectively,
the forms,
\bqn
\lb{3.33}
& & \left[\phi_{n}\right]^{-}  = - \frac{e^{-\sigma^{(1)}}}{\left(a^{2} -1\right)^{1/2}}
\frac{\partial V^{(1)}_{4}(\phi)}{\partial\phi},\\
\lb{3.34}
& & \frac{d{\rho}^{(1)}_{m}}{d\tau} + 3H_{u}\left(\rho^{(1)}_{m} + p^{(1)}_{m}\right) = 0,
\eqn
where $H_{u} \equiv \dot{a}_{u}/a_{u}$.

\subsubsection{Field Equations on  the surface $\Phi_{2} = 0$}

Following a similar procedure as what we did in the last sub-section, one can show that the Ricci 
tensor across the brane $\Phi_{2} = 0$ can be written as 
\bqn
\lb{3.41}
R_{ab} &=& R^{+}_{\;ab} H\left(\Phi_{2}\right) 
+  R^{-}_{\; ab} \left[1 - H\left(\Phi_{2}\right)\right]\nb\\
& & + R^{Im}_{\;ab}\delta\left(\Phi_{2}\right),
\eqn 	 
where $R^{+}_{\;ab}\; \left(R^{-}_{\; ab}\right)$ now is the Ricci tensor calculated in
the region $\Phi_{2} > 0\; \left(\Phi_{2} < 0\right)$, and $R^{Im}_{\;ab}$ denotes
the Ricci tensor calculated on the hypersurface $\Phi_{2} = 0$, which has the following
non-vanishing components,
\bqn
\lb{3.42}
R^{Im}_{\;tt} &=& L \left\{3\left[\omega_{l}\right]^{-} 
             - \left(b^{2} - 1\right)\left[\sigma_{l}\right]^{-}\right\},\nb\\
R^{Im}_{\;ty} &=& 3bL\left[\omega_{l}\right]^{-},\nb\\
R^{Im}_{\;yy} &=& L \left\{3b^{2}\left[\omega_{l}\right]^{-} 
             + \left(b^{2} - 1\right)\left[\sigma_{l}\right]^{-}\right\},\nb\\	     
R^{Im}_{\;ij} &=& L e^{2(\omega^{(2)} - \sigma^{(2)})} 
              \left(b^{2} - 1\right)\left[\omega_{l}\right]^{-}\delta_{ij},	        
\eqn
where $\omega_{l} \equiv l^{a}\omega_{,a}$ etc. On   the hypersurface $\Phi_{2} = 0$, 
the metric (\ref{3.1}) reduces to
\bq
\lb{3.43}
\left.ds^{2}_{5}\right|_{\Phi_{2} = 0} =  
            g^{(2)}_{\mu\nu}d\xi^{\mu}_{(2)}d\xi^{\nu}_{(2)}\nb\\
	    = d\eta^{2}  -a^{2}_{v}(\eta)d\Sigma^{2}_{0},
\eq
where $\xi^{\mu}_{(2)} \equiv \left\{\eta, \; x^{2}, \; x^{3},\; x^{4}\right\}$, and 
\bqn
\lb{3.44}
d\eta &\equiv& \epsilon_{\eta} \left(\frac{b^{2}-1}{b^{2}}\right)^{1/2} e^{\sigma^{(2)}} dt,\nb\\
a_{v}(\eta) &\equiv&   e^{\omega^{(2)}},
\eqn
with $ \epsilon_{\eta} = \pm 1$. Then, we find that
\bqn
\lb{3.45}
e^{(2)\; a}_{(\eta)} &\equiv& \frac{\partial x^{a}}{\partial\eta} 
           = {t}^{*}\left(\delta^{a}_{t} - \frac{1}{b}\delta^{a}_{y}\right),\nb\\
e^{(2)\; a}_{(i)} &\equiv& \frac{\partial x^{a}}{\partial{\xi^{i}_{(2)}}} 
           =  \delta^{a}_{i},\nb\\
\sqrt{\left|\frac{g^{(2)}_{4}}{g_{5}}\right|} &=& e^{-2\sigma^{(2)}},
\eqn	   
where ${t}^{*} \equiv dt/d\eta$. Hence, the field equations
of Eq.(\ref{2.16}) can be written as
\bqn
\lb{3.46a}
\left[\omega_{l}\right]^{-} &=& \frac{\kappa^{2}_{5}e^{-\sigma^{(2)}}}
      {3\left(b^{2} -1\right)^{1/2}} \left(\rho^{(2)}_{m} + V^{(2)}_{4}\right),\;\;\;\;\;\;\;\;\\
\lb{3.46b}
2\left[\omega_{l}\right]^{-} + \left[\sigma_{l}\right]^{-} &=& 
         \frac{\kappa^{2}_{5}e^{-\sigma^{(2)}}}
      {\left(b^{2} -1\right)^{1/2}}\left(V^{(2)}_{4} - p^{(2)}_{m}\right),\;\;
\eqn
where in writing the above equations we had assumed that $T^{(m,2)}_{\mu\nu}$ takes
the form, 
\bqn
\lb{3.47}
T^{(m,2)}_{\mu\nu} &\equiv& \left(\rho^{(2)}_{m} + p^{(2)}_{m}\right)w^{(2)}_{\mu}w^{(2)}_{\nu} 
             - p^{(2)}_{m} g^{(2)}_{\mu\nu},\nb\\
w^{(2)}_{\mu} &=& \delta^{\eta}_{\mu}.
\eqn 
Similarly, it can be shown that the Klein-Gordon equation  (\ref{2.18}) and the conservation law of
the matter fields (\ref{2.19}) on $\Sigma_{2}$ take, respectively,
the forms,
\bqn
\lb{3.48a}
& & \left[\phi_{l}\right]^{-}  = - \frac{e^{-\sigma^{(2)}}}{\left(b^{2} -1\right)^{1/2}}
\frac{\partial V^{(2)}_{4}(\phi)}{\partial\phi},\\
\lb{3.48b}
& & \frac{d{\rho}^{(2)}_{m}}{d\eta} + 3H_{v}\left(\rho^{(2)}_{m} + p^{(2)}_{m}\right) = 0,
\eqn
where $H_{v} \equiv  {a}^{*}_{v}/a_{v}$.

\section{Particular solutions for colliding  timelike 3-branes in the Einstein frame}

\renewcommand{\theequation}{4.\arabic{equation}}
\setcounter{equation}{0}

Choosing the potentials $V^{(I)}_{4}(\phi)$ on the two branes as
\bq
\lb{4.1a}
V^{(I)}_{4}(\phi) = V^{(I, 0)}_{4}e^{-\alpha\phi},
\eq
where $V^{(I, 0)}_{4}$'s and $\alpha$ are constants, and that the matter fields on each of the
two branes are dust fluids, i.e.,
\bq
\lb{4.1b}
p^{(I)}_{m} = 0,
\eq
we find a class of solutions,
which represents the collision of two timelike 3-branes and  is given by
\bqn
\lb{4.1}
\sigma &=& \left(\chi^{2} - \frac{1}{3}\right)\ln\left(X_{0} - X\right) + \sigma_{0},\nb\\
\omega &=&  \frac{1}{3} \ln\left(X_{0} - X\right) +  \omega_{0},\nb\\
\phi &=&  \frac{1}{\alpha} \ln\left(X_{0} - X\right) + \phi_{0},
\eqn
where $\chi \equiv  \kappa_{5}/(\sqrt{2}\alpha)$, $ A_{0},\; \sigma_{0},\; \omega_{0}$ and 
$\phi_{0}$ are arbitrary constants, and
\bqn
\lb{4.2}
X &=&  b\left(t-ay\right)H\left(\Phi_{1}\right)
        + a\left(t+by\right)H\left(\Phi_{2}\right)\nb\\
	 &=& \cases{(a+b)t, & IV,\cr
	 {a} \left(t+by\right), & III, \cr
	 {b}\left(t-ay\right), & II, \cr
	 0, & I.\cr} 
\eqn
The constants $a$ and $b$ are given by
\bqn
\lb{4.2a}
b\left(a^{2} - 1\right) &=& \frac{3\kappa^{2}_{5}V^{(1,0)}_{4}}{3\chi^{2} + 1},\nb\\
a\left(b^{2} - 1\right) &=& - \frac{3\kappa^{2}_{5}V^{(2,0)}_{4}}{3\chi^{2} + 1}.
\eqn
When $\alpha = \pm \infty$,    the solutions reduces to the 
ones studied previously \cite{TW08}. So, in the rest 
of this paper we shall consider only the case where $\alpha \not= \pm \infty$. 
Without loss of generality, we can always set $\sigma_{0} = \omega_{0} = \phi_{0} = 0$,
and assume that
\bq
\lb{4.2b}
X_{0} > 0.
\eq 


It can be shown that the field equations, Eqs.(\ref{3.9a}) and (\ref{3.9b}) [or
Eqs.(\ref{3.13a}), (\ref{3.13b}) and (\ref{3.16})], in Regions $I - IV$ are satisfied
identically for the above solutions. To study the singular behavior of the spacetime
in each of the four regions, we calculate the Ricci scalar, which in the present
case is given by
\bqn
\lb{4.2c}
R  &=& \kappa^{2}_{5}g^{ab} \phi_{,a} \phi_{,b}  \nb\\
&=& \frac{\kappa^{2}_{5} B}{\alpha^{2}\left(X_{0} - X\right)^{2(\chi^{2}+2/3)}},
\eqn
where $X$ is given by Eq.(\ref{4.2}), and
\bq
\lb{4.2d}
B  = \cases{ (a+b)^{2}, & IV, \cr
-a^{2}\left(b^{2} - 1\right), & III, \cr
-b^{2}\left(a^{2} - 1\right), & II, \cr
	 0, & I.\cr} 
\eq


On the 3-brane located on   $\Phi_{1} = 0$, the reduced metric takes
the form,
\bq
\lb{4.5a}
\left. ds^{2}_{5}\right|_{\Sigma_{1}} = d\tau^{2} - a^{2}_{u}(\tau)d^{2}\Sigma_{0},
\eq
where
\bq
\lb{4.5b}
a_{u}(\tau) = \cases{\left[\beta\left(\tau_{s} - \tau\right)\right]^{\frac{1}{3\chi^{2} + 2}},
  & $ \Phi_{2} > 0$,\cr
  X^{1/3}_{0}, & $ \Phi_{2} < 0$,\cr}
\eq
with
\bqn
\lb{4.5c}
\left.\Phi_{2}\right|_{\Phi_{1} = 0} &=&\frac{a+b}{a} t,\nb\\
\beta &\equiv& \frac{\left|a(a+b)\right|}{\left(a^{2} - 1\right)^{1/2}} \left(\chi^{2} + \frac{2}{3}\right),\nb\\
\tau_{s} &\equiv& \beta^{-1}X_{0}^{\chi^{2} + \frac{2}{3}}
\eqn
Note that in writing the above expressions, we had chosen $\epsilon_{\tau}
= {\mbox{sign}}(a+b)$. From Eqs.(\ref{3.31a}) and (\ref{3.31b}), on the other hand, we find that
\bqn
\lb{4.3}
\rho^{(1)}_{m} &=&  \frac{\rho^{(1, 0)}_{m}}{X_{0} - X^{(1)}(t)}\nb\\
&=&\cases{\left[\beta\left(\tau_{s} - \tau\right)\right]^{-\frac{3}{3\chi^{2} + 2}},
  & $ \Phi_{2} > 0$,\cr
  X^{-1}_{0}, & $ \Phi_{2} < 0$,\cr}
\eqn
where
\bqn
\lb{4.3a}
\rho^{(1, 0)}_{m} &\equiv& \frac{b\left(a^{2}-1\right)}
                      {\kappa^{2}_{5}}\left(\frac{2}{3} - \chi^{2}\right),\nb\\
X^{(1)}(t)  &\equiv& \left(a + b\right) t H\left(\Phi_{2}\right).		      
\eqn		 
From Eqs.(\ref{4.1}) and (\ref{4.2}) we also find that
\bq
\lb{4.3b}
\phi^{(1)}(\tau) =  
\cases
{\frac{1}{\alpha\left(3\chi^{2} + 2\right)} \ln\left[\beta\left(\tau_{s} - \tau\right)\right], 
  & $ \Phi_{2} > 0$,\cr
\frac{1}{\alpha}\ln{X_{0}}, & $ \Phi_{2} < 0$.\cr}
\eq 
		 

Similarly, on the 3-brane located on the hypersurface $\Phi_{2} = 0$,  
the reduced metric takes the form,
\bq
\lb{4.6a}
\left. ds^{2}_{5}\right|_{\Sigma_{2}} = d\eta^{2} - a^{2}_{v}(\eta)d^{2}\Sigma_{0},
\eq
where
\bq
\lb{4.6b}
a_{v}(\eta) = \cases{\left[\gamma\left(\eta_{s} - \eta\right)\right]^{\frac{1}{3\chi^{2} + 2}},
  & $ \Phi_{1} > 0$,\cr
  X^{1/3}_{0}, & $ \Phi_{1} < 0$,\cr}
\eq
with $\epsilon_{\eta} = {\mbox{sign}}(a+b)$, and 
\bqn
\lb{4.6c}
\left.\Phi_{1}\right|_{\Phi_{2} = 0} &=&\frac{a+b}{b} t,\nb\\
\gamma &\equiv& \frac{\left|b(a+b)\right|}{\left(b^{2} - 1\right)^{1/2}}
\left(\chi^{2} + \frac{2}{3}\right),\nb\\
\eta_{s} &\equiv& \gamma^{-1}X_{0}^{\chi^{2} + \frac{2}{3}}.
\eqn
The field equations (\ref{3.46a}) and (\ref{3.46b}), on the other hand, yield
\bqn
\lb{4.7}
\phi^{(2)}(\eta) &=&  \cases{\frac{1}{\alpha\left(3\chi^{2} + 2\right)}
\ln\left[\gamma\left(\eta_{s} - \eta\right)\right], & $ \Phi_{1} > 0$,\cr
\frac{1}{\alpha}\ln{X_{0}}, & $ \Phi_{1} < 0$,\cr},\nb\\
\rho^{(2)}_{m} &=&  \frac{\rho^{(2, 0)}_{m}}{X_{0} - X^{(2)}(t)}\nb\\
               &=& \cases
                   {\left[\gamma\left(\eta_{s}-\eta\right)\right]^{-\frac{3}{3\chi^{2} + 2}}, 
		   & $ \Phi_{1} > 0$,\cr
                   X^{-1}_{0},                                                                
		   & $ \Phi_{1} < 0$,\cr}
\eqn
where
\bqn
\lb{4.8}
\rho^{(2, 0)}_{m} &\equiv& - \frac{a\left(b^{2}-1\right)}
                      {\kappa^{2}_{5}}\left(\frac{2}{3} - \chi^{2}\right),\nb\\
X^{(2)}(t)  &\equiv& \left(a + b\right) t H\left(\Phi_{1}\right).		      
\eqn	

It is interesting to note that when $\chi^{2} = 2/3$, we have $\rho^{(I)}_{m} = 0,\; (I = 1, 2)$,
and the two 3-branes are supported only by the tensions $V^{(I)}_{4}(\phi)$, which are non-zero
for any finite value of $\alpha$ [Recall the conditions (\ref{3.4})]. It is also remarkable to
note that the presence of these two dust fluids is not essential to the singularity nature of the 
spacetime both in the bulk and on the branes. So, in the following we shall study the case with
$\chi^{2} = 2/3$ together with other cases.  

To study the above solutions further,  let us consider the  following cases separately:
(a) $\; a > 1, \; b > 1$; (b) $\; a > 1, \; b < - 1$; (c) $\; a < - 1, \; b > 1$;
and (d) $\; a < - 1, \; b < - 1$.

\subsection{$a > 1, \; b > 1$}

\begin{figure}
\includegraphics[width=\columnwidth]{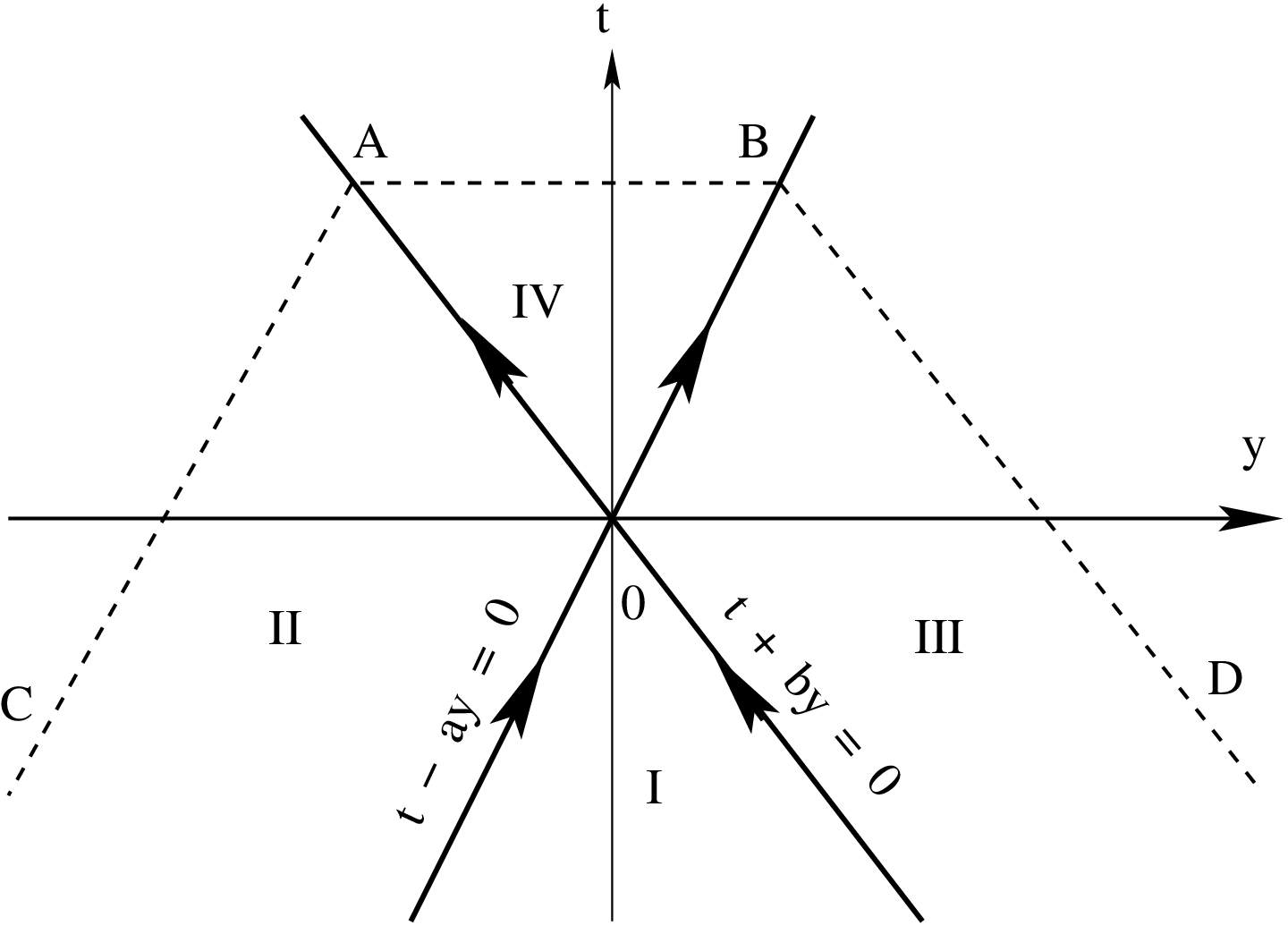}
\caption{The five-dimensional spacetime in the ($t, \; y$)-plane for 
$ a > 1,\; b  > 1$.  The two 3-branes  are moving along the hypersurfaces,
$\Sigma_{1}: t - ay = 0$ and $\Sigma_{2}: t + by = 0$. 
 $AB$ denotes the line $X_{0} = (a+b) t$, $AC$ the line $X_{0} = b(t -ay)$,
 and $BD$ the line $X_{0} = a(t +by)$. The spacetime is singular along these lines.
 The four regions, $I - IV$, are defined by Eq.(\ref{3.7}).}  
\label{fig1} 
\end{figure}

In this case, from Eq.(\ref{4.2a}) we find that
\bq
\lb{4.9}
 V^{(1)}_{4}(\phi) > 0, \;\;\;
 V^{(2)}_{4}(\phi) < 0,
\eq
while Eqs.(\ref{4.3}) and (\ref{4.7}) show that 
\bqn
\lb{4.10}
\rho^{(1)}_{m} &=& \cases{ \ge 0, & $\chi^{2} \le 2/3$,\cr
< 0, & $\chi^{2} > 2/3$,\cr},\nb\\
\rho^{(2)}_{m} &=& \cases{ \le 0, & $\chi^{2} \le 2/3$,\cr
> 0, & $\chi^{2} > 2/3$,\cr}.
\eqn
From Eq.(\ref{4.2c}) we can also see that  the spacetime is singular  along the 
line $X_{0} = (a+b) t$ in Region $IV$, the line $X_{0} = a(t + by)$
in Region $III$, and the line $X_{0} = b(t - ay)$ Region $II$, as shown by Fig. \ref{fig1}.

Before the collision $(t < 0)$, the scalar field is constant, $\phi^{(I)} = 
\phi_{1} \equiv (1/\alpha)\ln(X_{0})$, but both of the two potentials $V^{(1)}_{4}(\phi)$ 
and $V^{(2)}_{4}(\phi)$ are not zero, so do the dust energy densities $\rho^{(I)}_{m}$, 
except for  $\chi^{2} = 2/3$. In the case $\chi^{2} = 2/3$, the dust fluids disappear and 
the two branes are supported only by tensions, denoted by the two constant potential 
$V^{(1)}_{4}(\phi_{1})$ and $V^{(2)}_{4}(\phi_{1})$, which have the opposite signs, and
are quite similar to the case of Randall-Sundrum (RS) branes \cite{RS1}, except for that in the RS
model the two branes have $Z_{2}$ symmetry, while here we do not have. Before the collision, 
the spacetime on the two branes are flat,  that is,   the matter fields on the 3-brane do
not curve the 3-branes. However, it does curve the  spacetime outside the 3-branes. This is quite
similar to the so-called self-tuning mechanism of brane worlds \cite{cc}.

\begin{figure}
\includegraphics[width=\columnwidth]{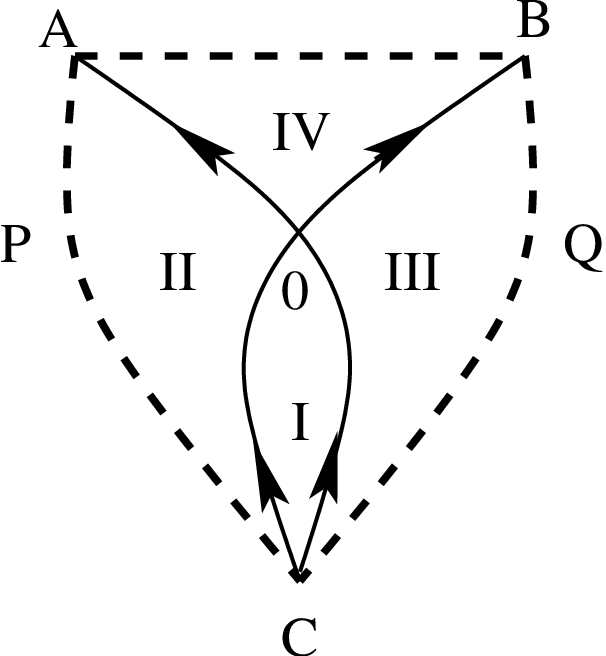}
\caption{The Penrose diagram for 
$ a >1,\;  b > 1$.  The spacetime is singular along the straight line $AB$ and the 
curved lines $APC$ and $BQC$.  }  
\label{fig1b} 
\end{figure}

After the collision, the two 3-branes focus each other and finally a spacetime singularity is 
developed at, respectively, $\tau = \tau_{s}$ and $\eta = \eta_{s}$.  The spacetime on the two branes
is homogeneous and isotropic, and is described, respectively, by Eqs.(\ref{4.5a})-(\ref{4.5b}) and
Eqs.(\ref{4.6a})-(\ref{4.6b}). The corresponding Penrose diagram is given by Fig. \ref{fig1b}.

\subsection{$a > 1, \; b < - 1$}

In this case, we find that  
\bqn
\lb{4.11}
& & V^{(1)}_{4}(\phi)  <  0, \;\;\;
 V^{(2)}_{4}(\phi) < 0, \nb\\
& & \rho^{(I)}_{m} = \cases{ \ge 0, & $ \chi^{2} \ge 2/3$,\cr
< 0, & $ \chi^{2} < 2/3$.\cr}
\eqn
Thus, unlike the last case, now both potentials $V^{(I)}_{4}(\phi)$ are negative, 
while the two dust energy densities always have the same sign. 

To study the solutions further in this case, we shall consider the two subcases, $a > |b| > 1$
and $  |b| > a > 1$, separately.

\subsubsection{$a > - b  > 1$}

\begin{figure}
\includegraphics[width=\columnwidth]{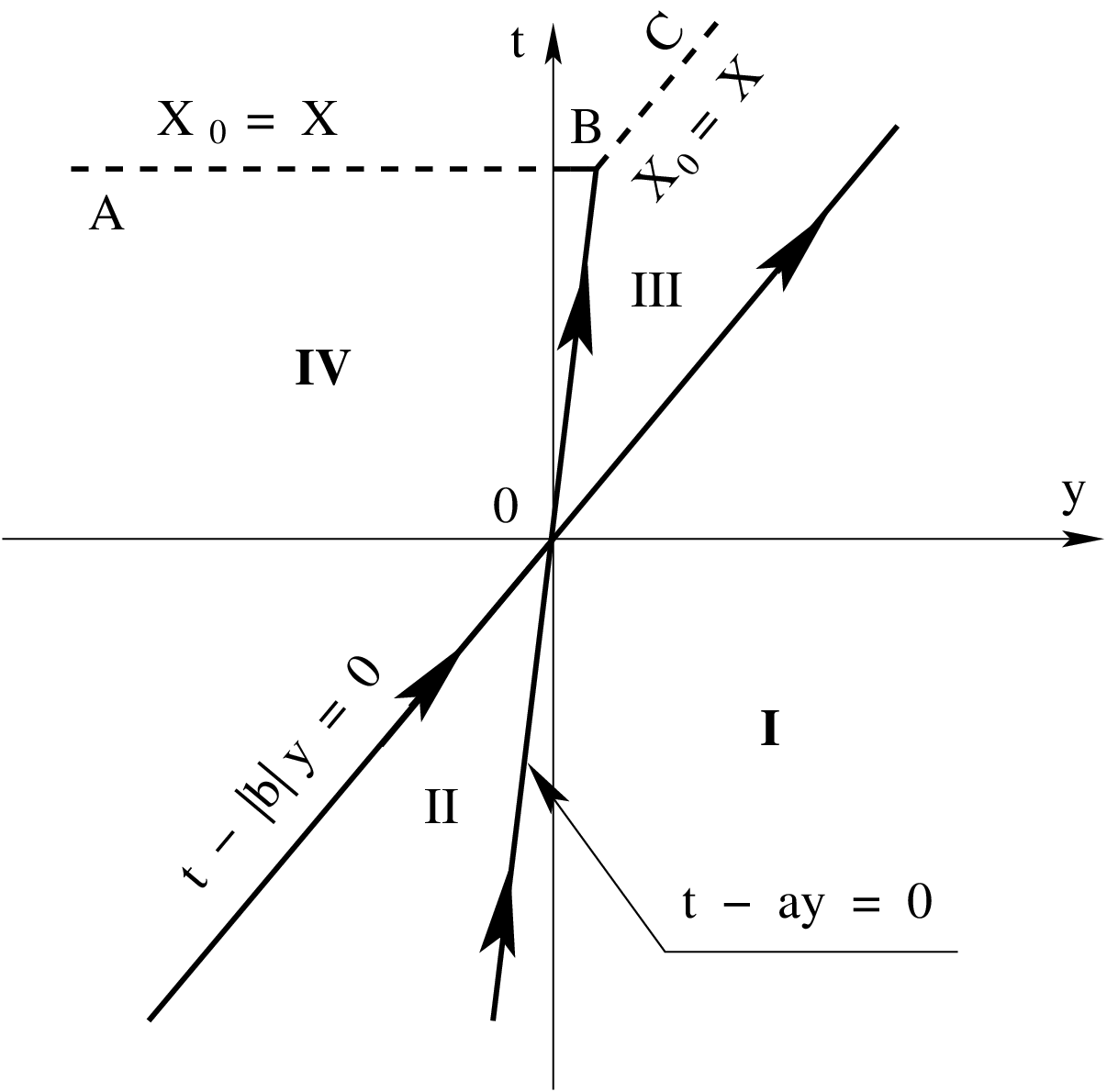}
\caption{The five-dimensional spacetime in the ($t, \; y$)-plane for 
$ a > - b > 1$.  The two 3-branes  are moving along the hypersurfaces,
$\Sigma_{1}: t - ay = 0$ and $\Sigma_{2}: t - |b|y = 0$. 
The spacetime is singular along the  line $AB$  in Region $IV$
and the line $BC$ in Region $III$. The spacetime is also singular on the 3-brane 
at the point $B$ where $\tau = \tau_{s}$. The four regions, $I - IV$, are 
defined by Eq.(\ref{3.7}).}  
\label{fig2} 
\end{figure}

When $a > -b > 1$, we have
\bqn
\lb{4.12}
\left.\Phi_{1}\right|_{\Phi_{2} = 0} &=& - \frac{a-|b|}{|b|} t
= \cases{< 0, & $t > 0$,\cr
> 0, & $t < 0$,\cr}\nb\\
\left.\Phi_{2}\right|_{\Phi_{1} = 0} &=&  \frac{a-|b|}{a} t
= \cases{> 0, & $t > 0$,\cr
< 0, & $t < 0$,\cr} \nb\\
 X_{0} - X &=& \cases{X_{0} - (a - |b|)t, & IV,\cr
	 X_{0} - a \left(t - |b|y\right), & III, \cr
	X_{0} +  |b|\left(t-ay\right), & II, \cr
	 0, & I.\cr} 
\eqn
Then, we find that the spacetime is singular along the line $X_{0} =  (a - |b|)t$ in Region $IV$, 
and the line $X_{0} = a \left(t - |b|y\right)$ in Region $III$, as shown in Fig. \ref{fig2}.

Before the collision $(t < 0)$, the scalar field   $\phi^{(1)}$ is constant on the 3-brane 
located on the hypersurface $\Sigma_{1}: t - ay = 0$, so does the  dust energy density 
$\rho^{(1)}_{m}$. In contrast, both the scalar field $\phi^{(2)}$ and the dust energy density 
$\rho^{(2)}_{m}$ are time-dependent on the 3-brane located on $\Sigma_{2}: t - |b|y = 0$, 
and the corresponding spacetime is described by Eqs.(\ref{4.6a}) and (\ref{4.6b}) with 
$\eta \le 0$. Note that along the hypersurface $\Sigma_{2}$, we have $\Phi_{1} > 0$
for $t < 0$, as shown by Eq.(\ref{4.12}). 

\begin{figure}
\includegraphics[width=\columnwidth]{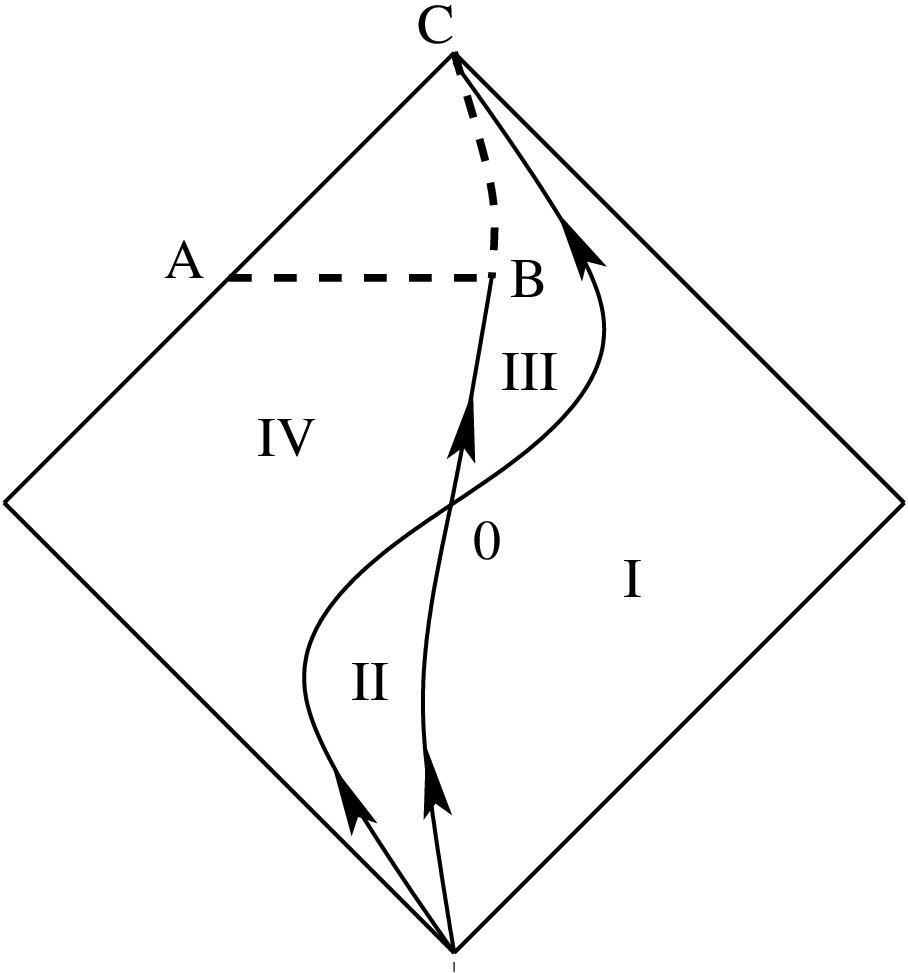}
\caption{The Penrose diagram for $ a > - b > 1$.  The spacetime is singular along the  
lines $AB$ and     $BC$.  }  
\label{fig2b} 
\end{figure}

After the collision, the  3-brane along $\Sigma_{2}$ transfers its energy to the one
along  $\Sigma_{1}$, so that its energy density $\rho^{(2)}_{m}$ and potential 
$V^{(2)}_{4}(\phi)$, as well as the scalar field $\phi^{(2)}$, become constant, 
while the energy density $\rho^{(1)}_{m}$ and  the scalar field $\phi^{(1)}$
become time-dependent. Because of the mutual focus of the two branes, a spacetime
singularity is finally developed at $\tau = \tau_{s}$, denoted by the point $B$ in Fig. 
\ref{fig2}. Afterwards, the spacetime becomes also singular along the line $X_{0}
= (a -|b|)t$  in Region $IV$ and the line $X_{0} = a(t -|b|y)$ in Region $III$.
It is interesting to note that these singularities are always formed, regardless of
the signs of $\rho^{(1)}_{m}$ and $\rho^{(2)}_{m}$. In fact, they are formed even
when $\rho^{(1)}_{m}(\chi^{2} = 2/3) = 0 = \rho^{(2)}_{m}(\chi^{2} = 2/3)$, as can be
seen from Eqs.(\ref{4.2c}), (\ref{4.5a}) and (\ref{4.5b}). This is because the scalar
field and the potentials $V^{(I)}_{4}(\phi)$ are still non-zero, and due the non-linear
interaction of the scalar field itself, spacetime singularities are still formed.
The corresponding Penrose diagram is given by Fig. \ref{fig2b}.

\subsubsection{$ - b > a > 1$}

\begin{figure}
\includegraphics[width=\columnwidth]{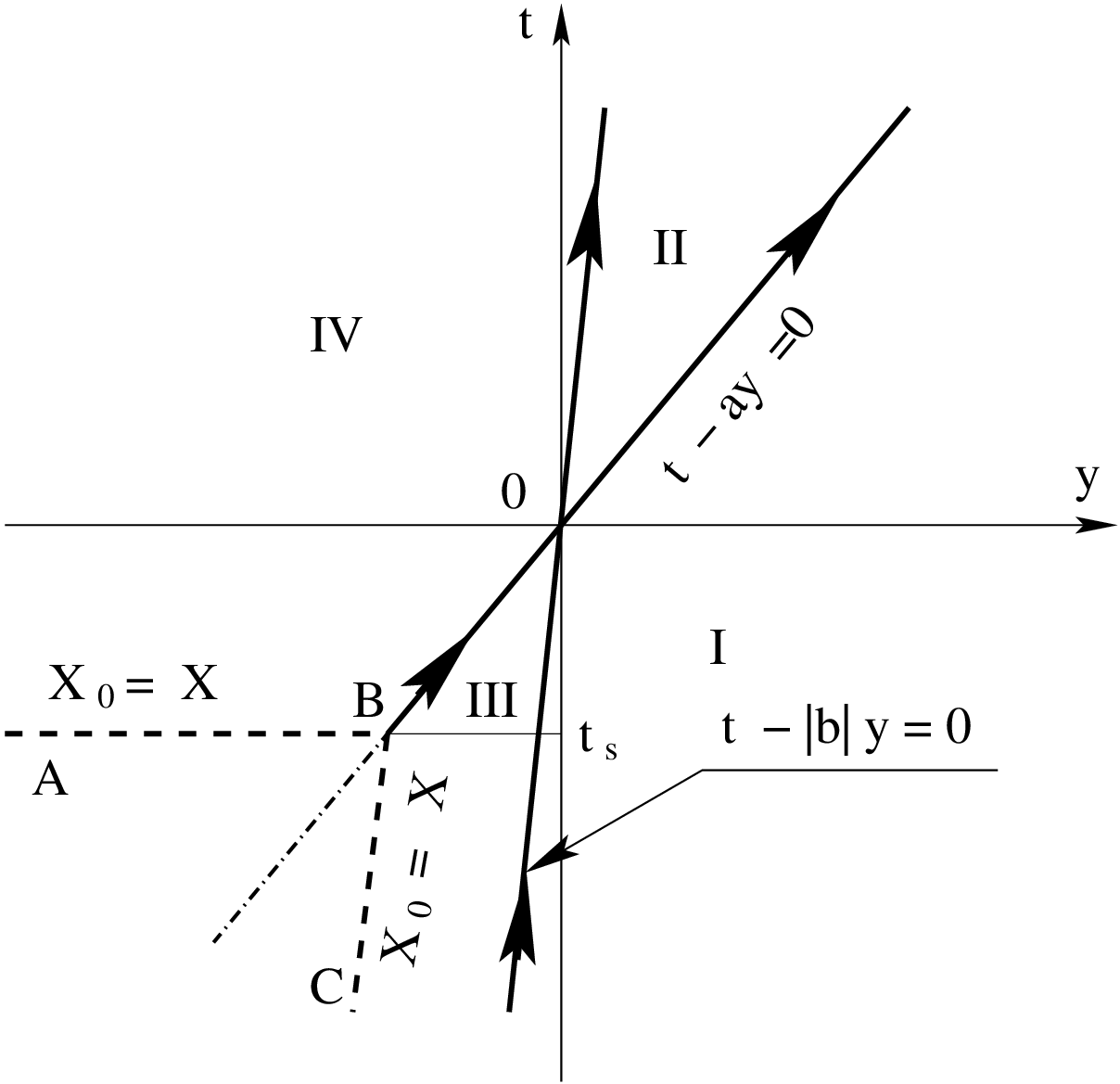}
\caption{The five-dimensional spacetime in the ($t, \; y$)-plane for 
$- b > a > 1$.  The two 3-branes  are moving along the hypersurfaces,
$\Sigma_{1}: t - ay = 0$ and $\Sigma_{2}: t - |b|y = 0$. 
The spacetime is singular along the  line $AB$  in Region $IV$
and the  line $BC$  in Region $III$. The spacetime is 
also singular on the 3-brane at the point $B$.}  
\label{fig3} 
\end{figure}

When $ - b > a > 1$, we have
\bqn
\lb{4.13}
\left.\Phi_{1}\right|_{\Phi_{2} = 0} &=& - \frac{|b| -a}{|b|} t
= \cases{> 0, & $t > 0$,\cr
< 0, & $t < 0$,\cr}\nb\\
\left.\Phi_{2}\right|_{\Phi_{1} = 0} &=&  - \frac{|b| -a}{a} t
= \cases{< 0, & $t > 0$,\cr
> 0, & $t < 0$,\cr} \nb\\
 X_{0} - X &=& \cases{X_{0} + (|b| -a)t, & IV,\cr
	 X_{0} - a \left(t - |b|y\right), & III, \cr
	X_{0} +  |b|\left(t-ay\right), & II, \cr
	 0, & I.\cr} 
\eqn
Then, we find that the spacetime is singular along the line $X_{0} =  - (|b| -a)t$ 
in Region $IV$ and the line $X_{0} =  (a - |b|)t$ in Region $III$, as shown in Fig. 
\ref{fig3}.

\begin{figure}
\includegraphics[width=\columnwidth]{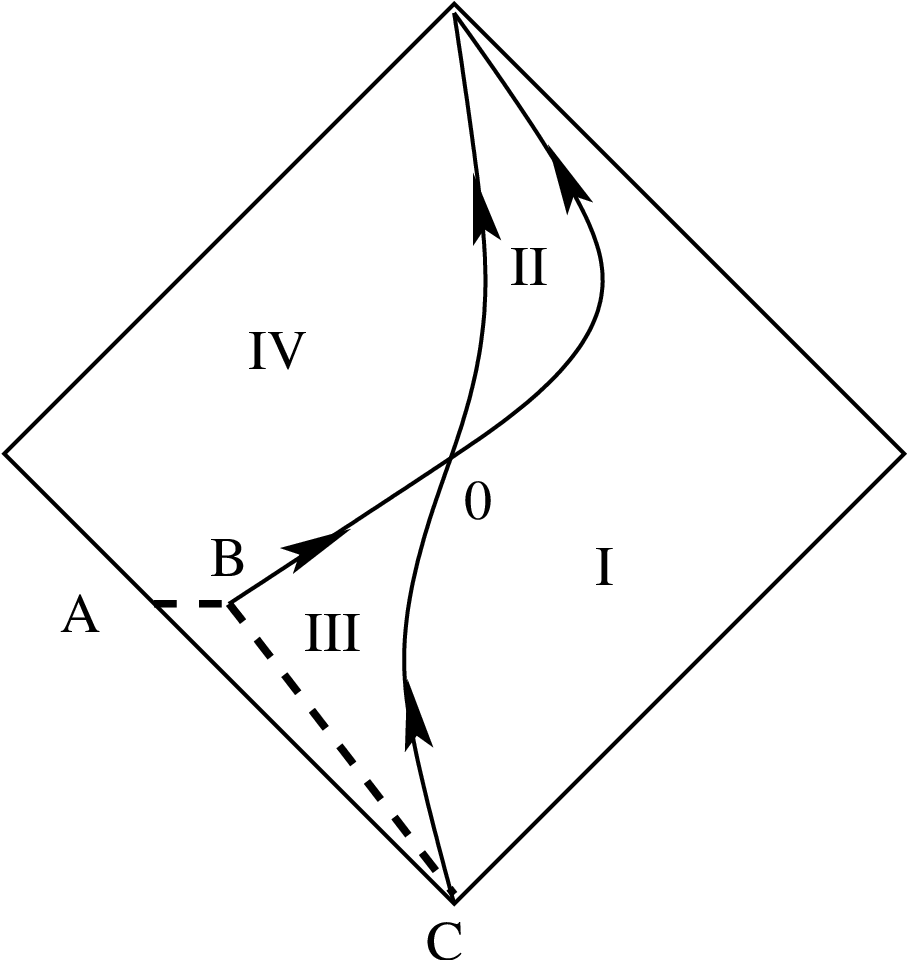}
\caption{The Penrose diagram for $ - b  > a > 1$.  The spacetime is singular along the  
lines $AB$ and   $BC$.  }  
\label{fig3b} 
\end{figure}

Unlike the last case, now the 3-brane on $\Sigma_{1}$ starts to expand at the
singular point $B$ where $\tau = \tau_{s}$, as shown in Fig. \ref{fig3}, and 
collides with the one on $\Sigma_{2}$ at the moment $\tau = 0 \; (t = 0)$. 
After the collision, its energy density $\rho^{(1)}_{m}$  the scalar field 
$\phi^{(2)}$ and the dust energy density $\rho^{(2)}_{m}$ on $\Sigma_{2}$ become 
time-dependent,  and the corresponding spacetime is described by Eqs.(\ref{4.6a}) 
and (\ref{4.6b}) with $\eta \in (0, - \infty)$.  The corresponding Penrose diagram 
is given by Fig. \ref{fig3b}.

\subsection{$a < - 1, \; b > 1$}

In this case, we find that  
\bqn
\lb{4.14}
& & V^{(I)}_{4}(\phi)  >  0, \nb\\
& & \rho^{(I)}_{m} = \cases{ \ge 0, & $ \chi^{2} \le 2/3$,\cr
< 0, & $ \chi^{2} > 2/3$,\cr}
\eqn
where $I = 1, \; 2$. Thus, in contrast to the last case, now both potentials $V^{(I)}_{4}(\phi)$ 
are positive, while the two dust energy densities always have the same sign.

\subsubsection{$ - a > b  > 1$}

\begin{figure}
\includegraphics[width=\columnwidth]{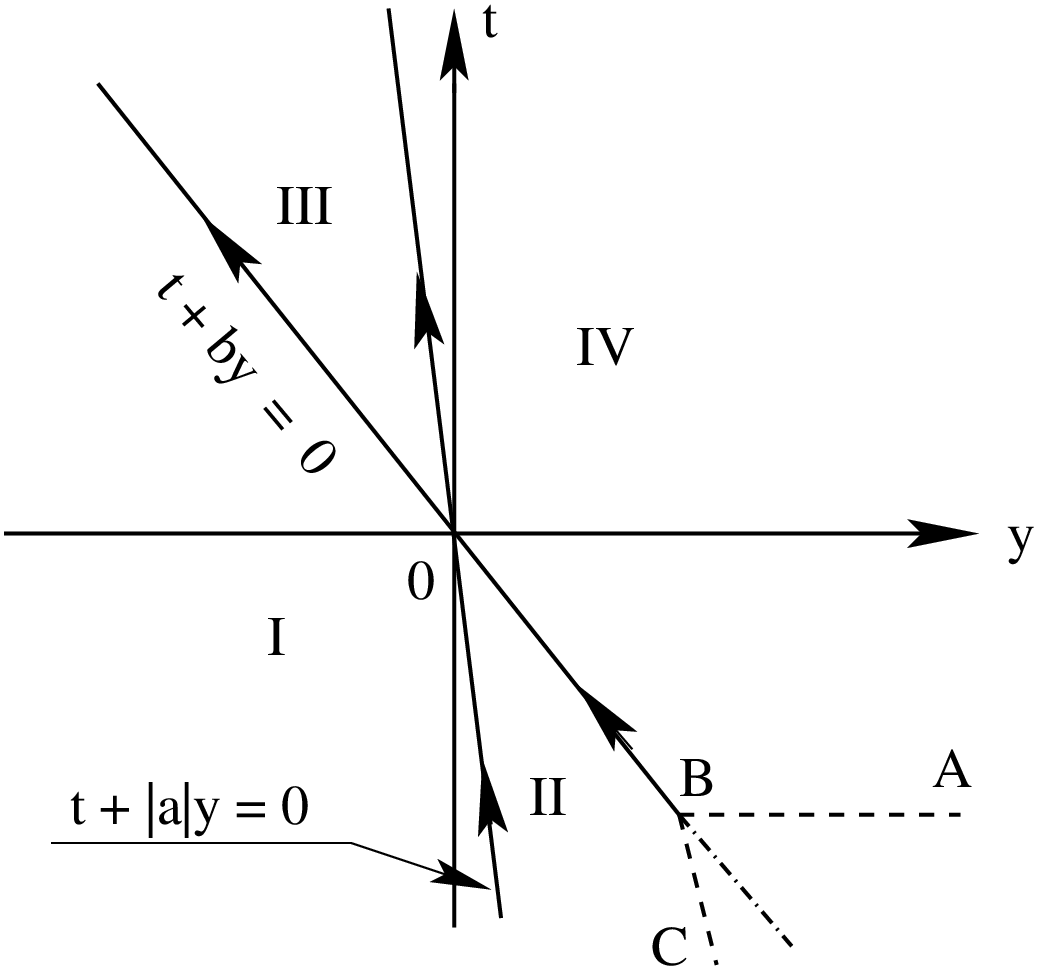}
\caption{The five-dimensional spacetime in the ($t, \; y$)-plane for 
$ - a > b > 1$.  The two 3-branes  are moving along the hypersurfaces,
$\Sigma_{1}: t + |a|y = 0$ and $\Sigma_{2}: t + by = 0$. 
The spacetime is singular along the  line $AB$  in Region $IV$
and the line $BC$ in Region $II$. The spacetime is also singular on the 3-brane 
at the point $B$ where $\eta = \eta_{s}$.}  
\label{fig4} 
\end{figure}

\begin{figure}
\includegraphics[width=\columnwidth]{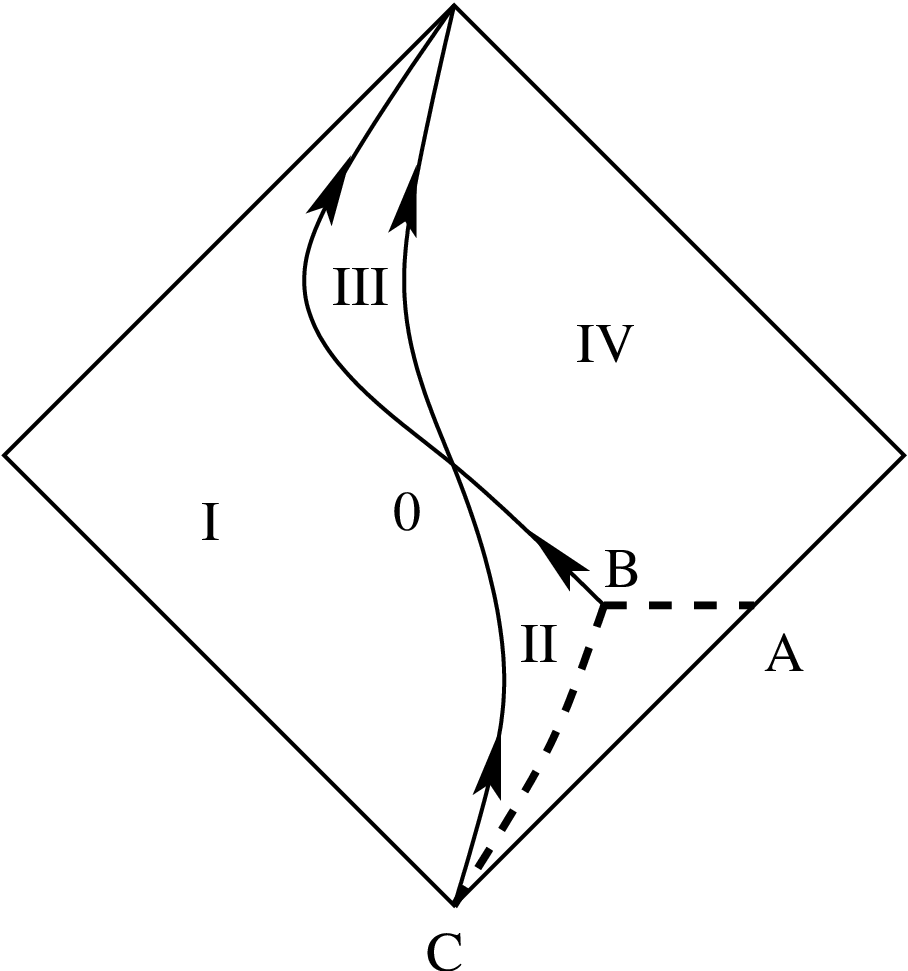}
\caption{The Penrose diagram for $ - a  > b > 1$.  The spacetime is singular along the  
lines $AB$ and   $BC$.  }  
\label{fig4b} 
\end{figure}

When $ - a > b > 1$, we have
\bqn
\lb{4.15}
\left.\Phi_{1}\right|_{\Phi_{2} = 0} &=& - \frac{|a|-b}{b} t
= \cases{< 0, & $t > 0$,\cr
> 0, & $t < 0$,\cr}\nb\\
\left.\Phi_{2}\right|_{\Phi_{1} = 0} &=&  \frac{|a|-b}{|a|} t
= \cases{> 0, & $t > 0$,\cr
< 0, & $t < 0$,\cr} \nb\\
 X_{0} - X &=& \cases{X_{0} + (|a| - b)t, & IV,\cr
	 X_{0} + |a| \left(t + by\right), & III, \cr
	X_{0} -  b\left(t+ |a|y\right), & II, \cr
	 0, & I.\cr} 
\eqn
Then, the spacetime is singular along the line $X_{0} = - (|a| - b)t$ in Region $IV$, 
and along the line $X_{0} = b \left(t + |a|y\right)$ in Region $II$, as shown in 
Fig. \ref{fig4}.  The corresponding Penrose diagram is given by Fig. \ref{fig4b}.

In this case, we also have
\bqn
\lb{4.16}
\phi^{(1)}(\tau) &=&  \cases{\frac{1}{\alpha\left(3\chi^{2} + 2\right)}
\ln\left[\beta\left(\tau_{s} - \tau\right)\right], & $ t > 0$,\cr
\frac{1}{\alpha}\ln{X_{0}}, & $ t < 0$,\cr}\nb\\
\phi^{(2)}(\eta) &=&  \cases{\frac{1}{\alpha}\ln{X_{0}}, & $ t > 0$,\cr
\frac{1}{\alpha\left(3\chi^{2} + 2\right)}
\ln\left[\gamma\left(\eta_{s} - \eta\right)\right], & $ t < 0$,\cr}\nb\\
\rho^{(1)}_{m} &=&\cases{\left[\beta\left(\tau_{s} 
- \tau\right)\right]^{-\frac{3}{3\chi^{2} + 2}},
  & $ t > 0$,\cr
  X^{-1}_{0}, & $ t < 0$,\cr}\nb\\
\rho^{(2)}_{m} &=&\cases{X^{-1}_{0}, & $ t > 0$,\cr
  \left[\gamma\left(\eta_{s} - \eta\right)\right]^{-\frac{3}{3\chi^{2} + 2}}.
  & $t < 0$.\cr} 
\eqn

\subsubsection{$b > - a > 1$}

\begin{figure}
\includegraphics[width=\columnwidth]{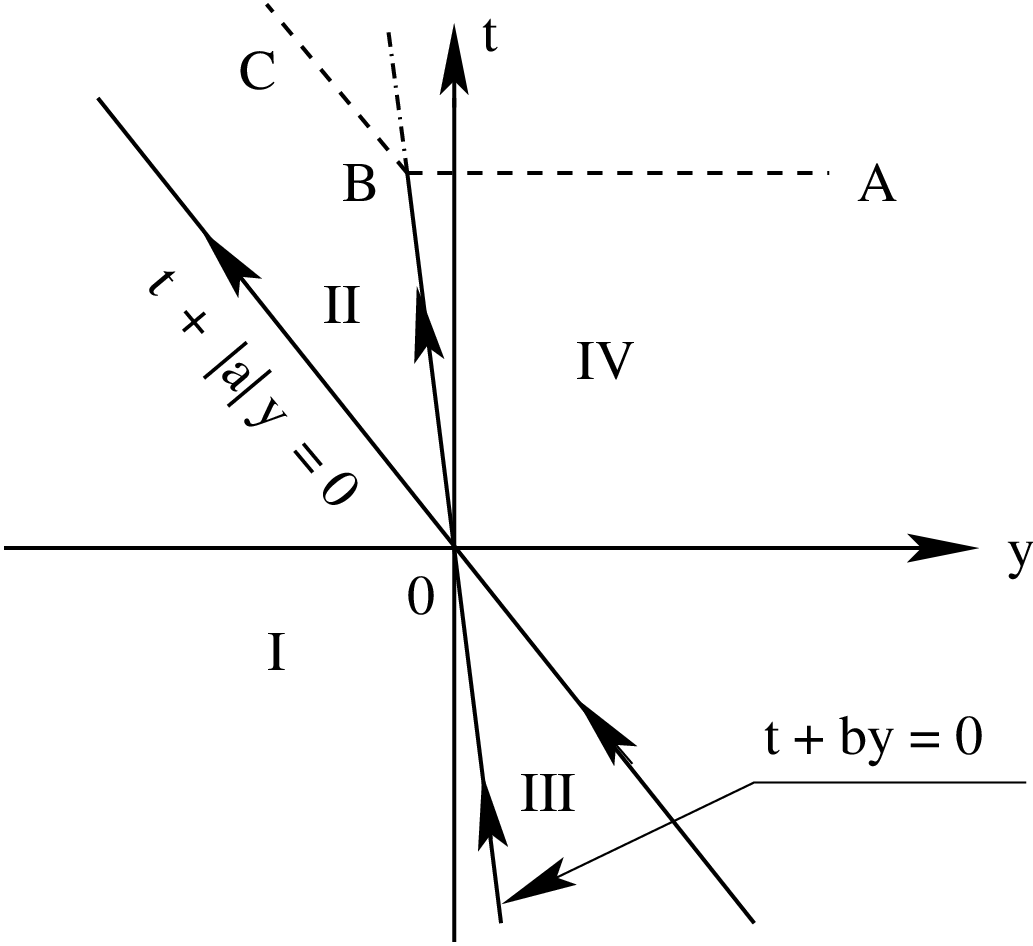}
\caption{The five-dimensional spacetime in the ($t, \; y$)-plane for 
$ b > - a > 1$.  The two 3-branes  are moving along the hypersurfaces,
$\Sigma_{1}: t + |a|y = 0$ and $\Sigma_{2}: t + by = 0$. 
The spacetime is singular along the  line $AB$  in Region $IV$
and the line $BC$ in Region $II$. The spacetime is also singular on the 3-brane 
at the point $B$ where $\eta = \eta_{s}$.}  
\label{fig5} 
\end{figure}

\begin{figure}
\includegraphics[width=\columnwidth]{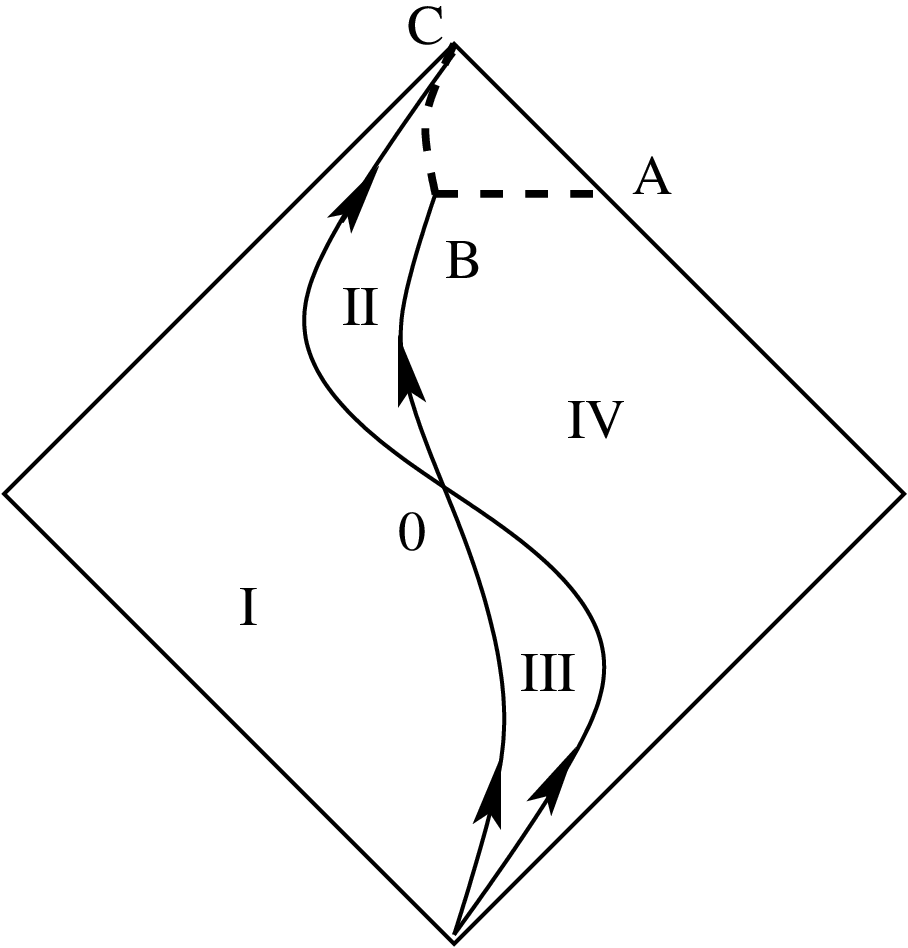}
\caption{The Penrose diagram for $ b  > -a > 1$.  The spacetime is singular along the  
lines $AB$ and   $BC$.  }  
\label{fig5b} 
\end{figure}

When $b > - a > 1$, we have
\bqn
\lb{4.17}
\left.\Phi_{1}\right|_{\Phi_{2} = 0} &=&  \frac{b - |a|}{b} t
= \cases{> 0, & $t > 0$,\cr
< 0, & $t < 0$,\cr}\nb\\
\left.\Phi_{2}\right|_{\Phi_{1} = 0} &=&  - \frac{b -|a|}{|a|} t
= \cases{< 0, & $t > 0$,\cr
> 0, & $t < 0$,\cr} \nb\\
 X_{0} - X &=& \cases{X_{0} - (b -|a|)t, & IV,\cr
	 X_{0} + |a| \left(t + by\right), & III, \cr
	X_{0} -  b\left(t+ |a|y\right), & II, \cr
	 0, & I.\cr} 
\eqn
We also have
\bqn
\lb{4.18}
\phi^{(1)}(\tau) &=&  \cases{\frac{1}{\alpha}\ln{X_{0}}, & $ t > 0$,\cr
\frac{1}{\alpha\left(3\chi^{2} + 2\right)}
\ln\left[\beta\left(\tau_{s} - \tau\right)\right], & $ t < 0$,\cr}\nb\\
\phi^{(2)}(\eta) &=&  \cases{\frac{1}{\alpha\left(3\chi^{2} + 2\right)}
\ln\left[\gamma\left(\eta_{s} - \eta\right)\right], & $ t > 0$,\cr
\frac{1}{\alpha}\ln{X_{0}}, & $ t < 0$,\cr}\nb\\
\rho^{(1)}_{m} &=&\cases{X^{-1}_{0}, & $ t > 0$,\cr
  \left[\beta\left(\tau_{s} - \tau\right)\right]^{-\frac{3}{3\chi^{2} + 2}},
  & $ t < 0$,\cr}\nb\\
\rho^{(2)}_{m} &=&\cases{
  \left[\gamma\left(\eta_{s} - \eta\right)\right]^{-\frac{3}{3\chi^{2} + 2}}.
  & $t > 0$,\cr
  X^{-1}_{0}, & $ t < 0$.\cr} 
\eqn
Then, the spacetime is singular along the line $X_{0} = (b-|a|)t$ in Region $IV$, 
and along the line $X_{0} = b \left(t + |a|y\right)$ in Region $II$, as shown 
in Fig. \ref{fig5}. The corresponding Penrose diagram is given by Fig. \ref{fig5b}.

\subsection{$a < - 1, \; b < - 1$}

\begin{figure}
\includegraphics[width=\columnwidth]{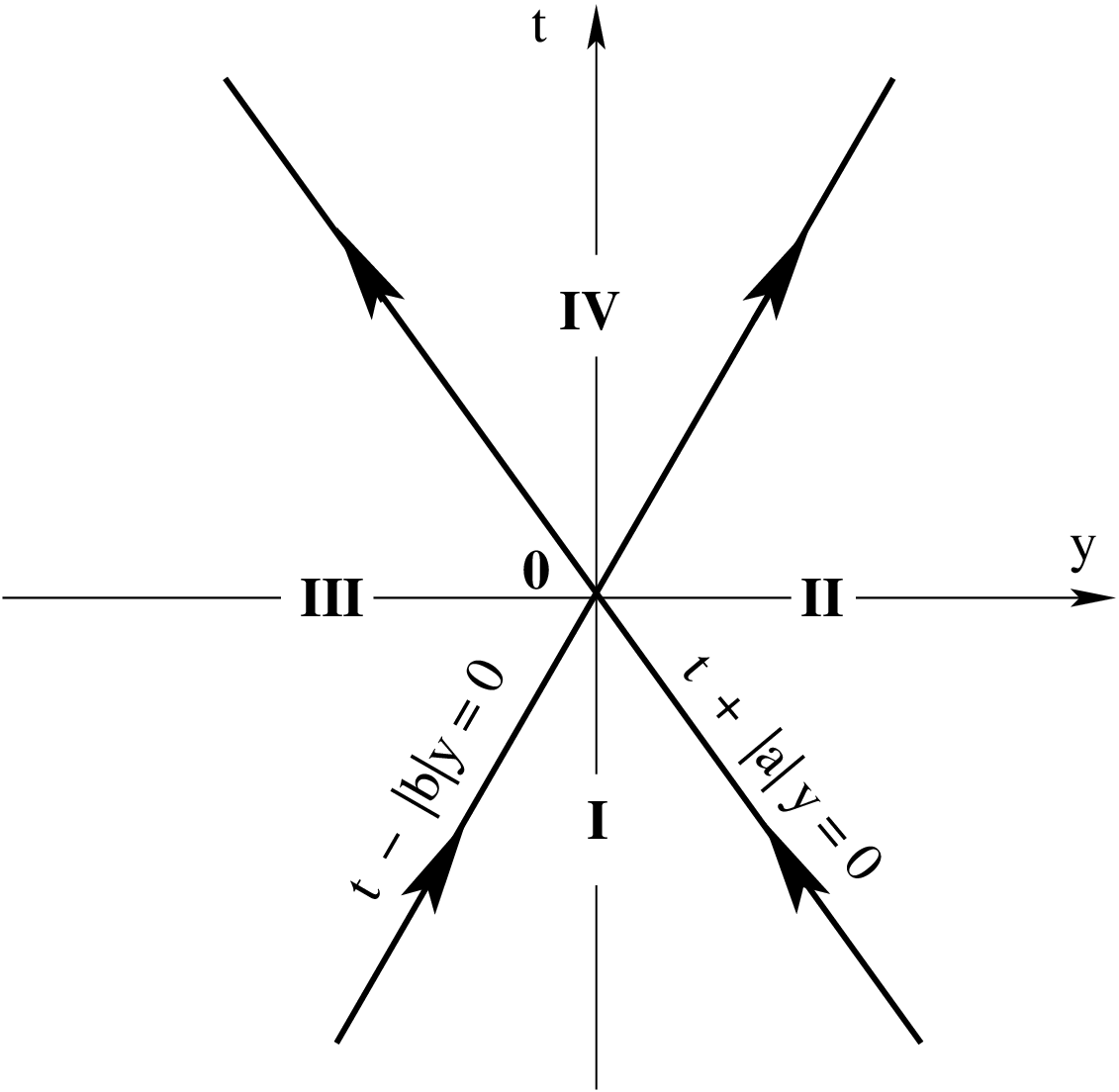}
\caption{The five-dimensional spacetime in the ($t, \; y$)-plane for 
$ a < - 1,\; b  < - 1$.  The two 3-branes  are moving along the hypersurfaces,
$\Sigma_{1}: t + |a| y = 0$ and $\Sigma_{2}: t - |b|y = 0$.   The spacetime is free
of any kind of spacetime singularities in the four regions, $I - IV$, as well
as on the two 3-branes. }  
\label{fig6} 
\end{figure}

\begin{figure}
\includegraphics[width=\columnwidth]{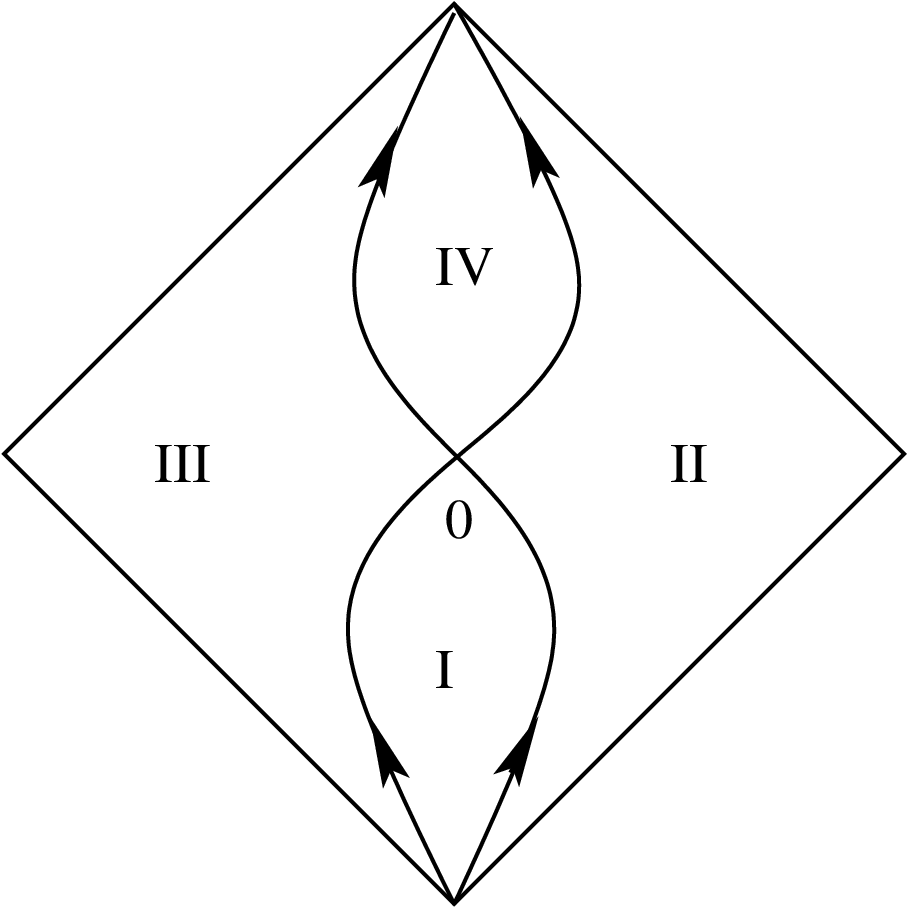}
\caption{The Penrose diagram for $ a < - 1,\; b  < - 1$.  The spacetime is non-singular
in all the regions.  }  
\label{fig6b} 
\end{figure}

In this case,  we have
\bqn
\lb{4.19}
 V^{(1)}_{4}(\phi) &<& 0, \;\;\;
 V^{(2)}_{4}(\phi) > 0,\nb\\
\rho^{(1)}_{m} &=& \cases{ \ge 0, & $\chi^{2} \ge 2/3$,\cr
< 0, & $\chi^{2} < 2/3$,\cr},\nb\\
\rho^{(2)}_{m} &=& \cases{ \ge 0, & $\chi^{2} \le 2/3$,\cr
> 0, & $\chi^{2} > 2/3$,\cr}
\eqn
and
\bqn
\lb{4.20}
\left.\Phi_{1}\right|_{\Phi_{2} = 0} &=&  \frac{|a| + |b|}{|b|} t
= \cases{> 0, & $t > 0$,\cr
< 0, & $t < 0$,\cr}\nb\\
\left.\Phi_{2}\right|_{\Phi_{1} = 0} &=& \frac{|a| + |b|}{|a|} t
= \cases{> 0, & $t > 0$,\cr
< 0, & $t < 0$,\cr} \nb\\
 X_{0} - X &=& \cases{X_{0} + (|a| + |b|)t, & IV,\cr
	 X_{0} + |a| \left(t  -|b|y\right), & III, \cr
	X_{0} + |b|\left(t + |a|y\right), & II, \cr
	 0, & I.\cr} 
\eqn
Then, we find that  
\bqn
\lb{4.21}
\phi^{(1)}(\tau) &=&  \cases{
\frac{1}{\alpha\left(3\chi^{2} + 2\right)}
\ln\left[\beta\left(\tau_{s} - \tau\right)\right], & $ t > 0$,\cr
\frac{1}{\alpha}\ln{X_{0}}, & $ t < 0$,\cr}\nb\\
\phi^{(2)}(\eta) &=&  \cases{\frac{1}{\alpha\left(3\chi^{2} + 2\right)}
\ln\left[\gamma\left(\eta_{s} - \eta\right)\right], & $ t > 0$,\cr
\frac{1}{\alpha}\ln{X_{0}}, & $ t < 0$,\cr}\nb\\
\rho^{(1)}_{m} &=&\cases{
  \left[\beta\left(\tau_{s} - \tau\right)\right]^{-\frac{3}{3\chi^{2} + 2}},
  & $ t > 0$,\cr
  X^{-1}_{0}, & $ t < 0$,\cr}\nb\\
\rho^{(2)}_{m} &=&\cases{
  \left[\gamma\left(\eta_{s} - \eta\right)\right]^{-\frac{3}{3\chi^{2} + 2}}.
  & $t > 0$,\cr
  X^{-1}_{0}, & $ t < 0$.\cr} 
\eqn
Note that in the present case, after the collision $t > 0$,  we have $\tau, \; \eta < 0$. 
Thus, in this case the spacetime is free of any kind singularity in all the four regions, 
as well as on the two branes, as shown in Fig. \ref{fig6}. The corresponding Penrose diagram 
is given by Fig. \ref{fig6b}.

It is interesting to note that when $\chi^{2} = 2/3$, the dust fluid on each of
the two 3-branes disappears, and the branes are supported only by the tensions,
where the brane along $\Sigma_{1}$ has a negative tension, while the one along
$\Sigma_{1}$ has a positive tension. It is also interesting to note that, when 
$\chi^{2} \not= 2/3$, both dust fluid are present, but they always have
opposite signs, that is, if one satisfies the energy conditions \cite{HE73}, the
other one must violate these conditions.

\section{Colliding    3-branes in the  5-dimensional string frame}

\renewcommand{\theequation}{5.\arabic{equation}}
\setcounter{equation}{0}

The spacetime singularity behavior in general can be quite different
in the two frames, due to the conformal transformations of Eq.(\ref{2.7}), 
which are  often singular.  The 5-dimensional spacetime in the string frame is 
given by
\bqn
\lb{5.1}
d^{2}\hat{s}_{5} &\equiv& \gamma_{ab} dx^{a}dx^{b}\nb\\
&=&  e^{2\hat{\sigma}(t,y)}\left(dt^{2} - dy^{2}\right)
- e^{2\hat{\omega}(t,y)}d\Sigma^{2}_{0},
\eqn
where $d\Sigma^{2}_{0}$ is given in Eq.(\ref{3.1}), and
\bqn
\lb{5.2}
\hat{\sigma}(t,y) &\equiv& \left(\chi^{2} - \epsilon  \sqrt{\frac{5}{12}}\; \chi - 
\frac{1}{3}\right)\ln\left(X_{0} - X\right),\nb\\
\hat{\omega}(t,y)  &\equiv& \left(\frac{1}{3}  - \epsilon \sqrt{\frac{5}{12}}\; 
\chi\right)\ln\left(X_{0} - X\right),\nb\\
\hat{\phi}(t,y) &\equiv& \left(X_{0} - X\right)^{\epsilon  \sqrt{\frac{3}{20}}\; \chi}, 
\eqn
where $\epsilon = \pm 1$.  

\subsubsection{The Spacetime Singularities in Regions $I - IV$}

To study the spacetime singularities in Regions $I - IV$, let us consider the quantity, 
\bq
\lb{5.3}
\hat{\phi}_{,a} \hat{\phi}^{,a} = \frac{3\chi^{2}B}
{20\left(X_{0} - X\right)^{\frac{4}{5} + \left(\sqrt{\frac{8}{15}} 
- \epsilon  \sqrt{2}\; \chi\right)^{2}}},
\eq
where $B$ is given by Eq.(\ref{4.2d}).  Comparing the above expression with
Eq.(\ref{4.2c}), we find that  the spacetime in Regions $I - IV$
is singular in the string frame whenever it  is singular in the Einstein frame, 
although the strength of the singularity is different, as can be seen clearly
from the following expression,
\bq
\lb{5.5}
20 \frac{\hat{\phi}_{,a} \hat{\phi}^{,a}}{R} = 
\left(X_{0} - X\right)^{\epsilon\chi\sqrt{\frac{64}{15}}},
\eq
In particular, if $\epsilon\alpha > 0$ the singularity in the Einstein frame
is stronger, and if $\epsilon\alpha < 0$ it is the other way around. 
 
\subsubsection{The Spacetime on the 3-brane $t = ay$}

On the hypersurface $t = ay$, the metric (\ref{5.1}) reduces to
\bq
\lb{5.6a}
\left. d^{2}\hat{s}_{5}\right|_{t = ay} =   d\hat{\tau}^{2}
- a^{2}_{u}\left(\hat{\tau}\right) d\Sigma^{2}_{0},
\eq
where
\bqn
\lb{5.7a}
a_{u}(\hat{\tau}) &=& \cases{a_{0}\left(\hat{\tau}_{s} 
- \hat{\tau}\right)^{\Delta}, & $ \Phi_{2} > 0$,\cr
a_{0} \hat{\tau}_{s}^{\Delta}, & $ \Phi_{2} < 0$,\cr}\nb\\
\hat{\phi}^{(1)}(\hat{\tau}) &=&  \cases{\left[\hat{\beta}\left(\hat{\tau}_{s} 
- \hat{\tau}\right)\right]^{\epsilon\sqrt{\frac{3}{20}}\frac{\chi}{\delta}},
  & $ \Phi_{2} > 0$,\cr
  \left(\hat{\beta}\hat{\tau}_{s}\right)^{\epsilon\sqrt{\frac{3}{20}}\frac{\chi}{\delta}}, 
  & $ \Phi_{2} < 0$,\cr}
\eqn
with
\bqn
\lb{5.8a}
X_{0} - X^{(1)} &=& \cases{\left[\hat{\beta}\left(\hat{\tau}_{s} 
- \hat{\tau}\right)\right]^{\frac{1}{\delta}}, & $\Phi_{2} > 0$,\cr
X_{0}, & $\Phi_{2} < 0$,\cr}\nb\\
\left.\Phi_{2}\right|_{\Phi_{1} = 0} &=&\frac{a+b}{a} t,\;\;\;\;\;
\hat{\beta} \equiv \frac{\left|a(a+b)\right|}{\sqrt{a^{2} -1}} \delta,\nb\\
\hat{\tau}_{s} &\equiv& \hat{\beta}^{-1}X_{0}^{\delta}, \;\;\;
a_{0} \equiv \hat{\beta}^{\Delta}.\nb\\
\delta &\equiv& \left(\sqrt{\frac{5}{48}} - \epsilon \chi\right)^{2} + \frac{9}{16} > 0,\nb\\
\Delta &\equiv& \frac{1}{\delta}\left(\frac{1}{3} - \epsilon\chi\sqrt{\frac{5}{12}}\right). 
\eqn
Note that in writing the above expressions, we had chosen $\epsilon_{\hat{\tau}}
= {\mbox{sign}}(a+b)$. To study the spacetime singularity on the brane, we calculate the Ricci
scalar, which now is given by
\bq
\lb{5.9a}
R^{(4)\lambda}_{u\;\;\;\;\lambda}  =  \frac{3\Delta\left(2-\Delta\right)}{2a_{0}\left(\hat{\tau}_{s} 
- \hat{\tau}\right)^{\Delta + 2}},
\eq
where
\bqn
\lb{5.10a}
\Delta + 2 &=&  \frac{1}{\delta}\left[2\left(\epsilon\chi - \sqrt{\frac{15}{64}}\right)^{2}
 + \frac{115}{96}\right] > 0,\nb\\
 \Delta - 2 &=&  - \frac{1}{\delta}\left[2\left(\epsilon\chi - \sqrt{\frac{5}{48}}\right)^{2}
 + \frac{19}{24}\right] < 0.
\eqn

\subsubsection{The Spacetime on the 3-brane $t = -by$}

Similarly, on the 3-brane located on the hypersurface $\Phi_{2} = 0$,  
 the metric (\ref{5.1}) reduces to
\bq
\lb{5.6b}
\left. d^{2}\hat{s}_{5}\right|_{t = -by} =   d\hat{\eta}^{2}
- a^{2}_{v}\left(\hat{\eta}\right) d\Sigma^{2}_{0},
\eq
where 
where
\bqn
\lb{5.7b}
a_{v}(\hat{\eta}) &=& \cases{a_{0}\left(\hat{\eta}_{s} 
- \hat{\eta}\right)^{\Delta}, & $ \Phi_{1} > 0$,\cr
a_{0} \hat{\eta}_{s}^{\Delta}, & $ \Phi_{1} < 0$,\cr}\nb\\
\hat{\phi}^{(2)}(\hat{\eta}) &=&  \cases{\left[\hat{\gamma}\left(\hat{\eta}_{s} 
- \hat{\eta}\right)\right]^{\epsilon\sqrt{\frac{3}{20}}\frac{\chi}{\delta}},
  & $ \Phi_{1} > 0$,\cr
  \left(\hat{\gamma}\hat{\eta}_{s}\right)^{\epsilon\sqrt{\frac{3}{20}}\frac{\chi}{\delta}}, 
  & $ \Phi_{1} < 0$,\cr}
\eqn
with
\bqn
\lb{5.8b}
X_{0} - X^{(2)} &=& \cases{\left[\hat{\gamma}\left(\hat{\eta}_{s} 
- \hat{\eta}\right)\right]^{\frac{1}{\delta}}, & $\Phi_{1} > 0$,\cr
X_{0}, & $\Phi_{1} < 0$,\cr}\nb\\
\left.\Phi_{1}\right|_{\Phi_{2} = 0} &=&\frac{a+b}{b} t,\;\;\;
\hat{\gamma} \equiv \frac{\left|a(a+b)\right|}{\sqrt{b^{2} -1}} \delta,\nb\\
\hat{\eta}_{s} &\equiv& \hat{\gamma}^{-1}X_{0}^{\delta},
\eqn
but now we have $a_{0} \equiv \hat{\gamma}^{\Delta}$ and $\epsilon_{\hat{\eta}}
= {\mbox{sign}}(a+b)$. For the metric (\ref{5.6b}),  we also find  that
\bq
\lb{5.9b}
R^{(4) \lambda}_{v\;\;\;\;\lambda}  =  \frac{3\Delta\left(2-\Delta\right)}{2a_{0}\left(\hat{\eta}_{s} 
- \hat{\eta}\right)^{\Delta + 2}}.
\eq
From Eqs.(\ref{5.9a}) and (\ref{5.9b}) we can see that the spacetime on each of the branes
is not singular when $\Delta = 0$ or $\chi = \epsilon\sqrt{\frac{4}{15}}$. As a matter of fact,
in this case the spacetime on each of the two branes is flat.  Thus, in the following we need to
consider only the case $\chi \not= \epsilon\sqrt{\frac{4}{15}}$.

From Eqs.(\ref{5.7a})-(\ref{5.10a}) and Eqs.(\ref{5.7b})-(\ref{5.9b}), it can be shown that the 
spacetime singularities on each of the two branes are similar to these in the Einstein frame.
For example, for the case $ a > 1, \; b > 1$, it is singular at $\hat{\tau} = \hat{\tau}_{s}$
and $\hat{\eta} = \hat{\eta}_{s}$, which correspond to, respectively, the point $A$ and $B$ in
Fig. \ref{fig1b}. Similarly, the spacetime is free from any kind of singularities for the case
$a <- 1, \; b < - 1$, and the corresponding Penrose diagram is also given by Fig. \ref{fig6b}.

\section{Colliding 3-branes in  the 10-dimensional Spacrtimes}

\renewcommand{\theequation}{6.\arabic{equation}}
\setcounter{equation}{0}

Lifting the metric to 10-dimensions,  it is given by Eq.(\ref{3.1}),
which can be cast in the form,
\bqn
\lb{6.1}
d^{2}\hat{s}_{10} &\equiv& \gamma_{ab} dx^{a}dx^{b} 
+ \hat{\phi}^{2}\left(x^{c}\right) 
\hat{\gamma}_{ij}\left(z^{k}\right)dz^{i} dz^{j} \nb\\
&=&  e^{2\hat{\sigma}(t,y)}\left(dt^{2} - dy^{2}\right)
- e^{2\hat{\omega}(t,y)}d\Sigma^{2}_{0} \nb\\
& & - \hat{\phi}^{2}\left(t, y\right) d\Sigma_{z}^{2},
\eqn
where  $\hat{\sigma},\; \hat{\omega}$ and $\hat{\phi}$ are given by 
Eq.(\ref{5.2}), and $d\Sigma_{z}^{2} \equiv - \sum^{5}_{i,j =1}{
\hat{\gamma}_{ij}\left(z^{k}\right)dz^{i} dz^{j}}$. Then, it can be
shown that the spacetime in Regions $I - IV$ is vacuum,
\bq
\lb{6.2}
R^{(A)}_{AB} = 0,
\eq
where $A = I, ..., IV$, as it is expected. To study the singular behavior of the spacetime
in these regions, we calculate the Kretschmann scalar, which in the present case is given by
\bqn
\lb{6.3}
I_{10} &\equiv& R_{ABCD}  R^{ABCD}\nb\\
&=& \frac{B^{2} I^{(0)}_{10}}{\left(X_{0} - X\right)^{\left(2\chi 
- \epsilon \sqrt{\frac{5}{12}}\right)^{2} + \frac{9}{4}}},
\eqn
where $B$ is given by Eq.(\ref{4.2d}), and
\bqn
\lb{6.4}
I_{10}^{(0)} &\equiv& \frac{1}{45}\left[\left(720\chi^{6}
+ 1287\chi^{4} + 200\chi^{2} +40\right)\right.\nb\\
& & \left. - 312\epsilon\sqrt{\frac{5}{3}}\chi^{3}\left(2
+ 3\chi^{2}\right)\right].
\eqn
It can be shown that  $I_{10}^{(0)}$ is non-zero for any given $\chi$. Then, comparing the
expression of Eq.(\ref{6.3}) with Eq.(\ref{4.2c}), we find that the lifted  10-dimensional
spacetime has a similar singular behavior as that in the 5-dimensional spacetime in the 
Einstein frame. In particular, it is also singular on the hypersurface $X_{0} - X =0$.


On the hypersurface $t = ay$, the metric (\ref{6.1}) reduces to
\bq
\lb{6.5}
\left. d^{2}\hat{s}_{5}\right|_{t = ay} =   d\hat{\tau}^{2}
- a^{2}_{u}\left(\hat{\tau}\right) d\Sigma^{2}_{0}
-  b^{2}_{u}\left(\hat{\tau}\right) d\Sigma_{z}^{2},
\eq
where $a_{u}\left(\hat{\tau}\right)$ and $ b_{u}\left(\hat{\tau}\right) \equiv
{\hat{\phi}^{(1)}\left(\hat{\tau}\right)}$ are given by Eqs.(\ref{5.7a}) and 
(\ref{5.8a}). On the 8-brane, the Einstein tensor has distribution given by
Eqs.(\ref{a.6}) and (\ref{a.7}). Inserting Eq.(\ref{5.2}) into Eq.(\ref{5.8a}),
and noticing that $\hat{\psi} \equiv \ln\left(\hat{\phi}\right)$, we find
\bqn
\lb{6.6}
\hat{\rho}_{u}   &=& \frac{b\left(a^{2} - 1\right)}{\left[X_{0} - X^{(1)}(t)\right]^\mu},\nb\\ 
\hat{p}^{Z}_{u} &=&  -\frac{b\left(a^{2} - 1\right)}{\left[X_{0} - X^{(1)}(t)\right]^\mu}
                    \left[\left(\chi - \epsilon\sqrt{\frac{4}{15}}\right)^{2} 
		    + \frac{2}{5}\right],\nb\\
\hat{p}^{X}_{u} &=&  -\frac{b\left(a^{2} - 1\right)}{\left[X_{0} - X^{(1)}(t)\right]^\mu}
                    \left(\chi^{2} + \frac{1}{3}\right), 		    
\eqn
where $X^{(1)}(t)$ is given by Eq.(\ref{5.8a}), and 
\bq
\lb{6.7}
\mu \equiv 2\left(\chi - \epsilon\sqrt{\frac{5}{48}}\right)^{2} + \frac{1}{8}.
\eq 
Clearly, whenever $X_{0} -  X^{(1)}(t) = 0$, the spacetime on the 8-brane is
singular.


On the hypersurface $t = -by$, the metric (\ref{6.1}) reduces to
\bq
\lb{6.5a}
\left. d^{2}\hat{s}_{5}\right|_{t = -by} =   d\hat{\eta}^{2}
- a^{2}_{v}\left(\hat{\eta}\right) d\Sigma^{2}_{0}
-  b^{2}_{v}\left(\hat{\eta}\right) d\Sigma_{z}^{2},
\eq
where $a_{v}\left(\hat{\eta}\right)$ and $ b_{v}\left(\hat{\eta}\right) \equiv
{\hat{\phi}^{(2)}\left(\hat{\eta}\right)}$ are given by Eqs.(\ref{5.7b}) and 
(\ref{5.8b}). On this 8-brane, the Einstein tensor has distribution given by
Eqs.(\ref{a.6a}) and (\ref{a.7a}), which in the present case yield,
\bqn
\lb{6.6a}
\hat{\rho}_{v}   &=& \frac{a\left(b^{2} - 1\right)}{\left[X_{0} - X^{(2)}(t)\right]^\mu},\nb\\ 
\hat{p}^{Z}_{v} &=&  -\frac{a\left(b^{2} - 1\right)}{\left[X_{0} - X^{(2)}(t)\right]^\mu}
                    \left[\left(\chi - \epsilon\sqrt{\frac{4}{15}}\right)^{2} 
		    + \frac{2}{5}\right],\nb\\
\hat{p}^{X}_{v} &=&  -\frac{a\left(b^{2} - 1\right)}{\left[X_{0} - X^{(2)}(t)\right]^\mu}
                    \left(\chi^{2} + \frac{1}{3}\right), 		    
\eqn
where $X^{(2)}(t)$ is given by Eq.(\ref{5.8b}).   Thus,  the  spacetime on this 8-brane is
also singular  whenever $X_{0} -  X^{(2)}(t) = 0$.

When $a > 1$ and $b > 1$, from Eqs.(\ref{6.6}) and (\ref{6.6a})	it can be shown that 
both of the weak and dominant energy conditions \cite{HE73} are satisfied by the matter
fields on the two 8-branes, provided that
\bq
\lb{6.8}
\cases{\sqrt{\frac{4}{15}} - \sqrt{\frac{3}{5}}  \le \chi \le \sqrt{\frac{2}{3}}, &
$\epsilon = +1$, \cr
- \sqrt{\frac{2}{3}} \le \chi \le  \sqrt{\frac{3}{5}}-\sqrt{\frac{4}{15}}, &
$\epsilon = -1$, \cr}
\eq
but the strong energy condition is always violated. 
When $a > 1$ and $b < - 1$, the matter field on the 8-brane $\Phi_{1} = 0$ violates all the 
three energy conditions, while the one on the 8-brane $\Phi_{2} = 0$ satisfies the weak and
dominant energy conditions, provided that the conditions (\ref{6.8}) holds, but violates the 
strong one. When $a  < -1$ and $b > 1$, it is the other way around, that is, the matter field 
on the 8-brane $\Phi_{1} = 0$ satisfies the weak and dominant energy conditions, provided 
that the conditions (\ref{6.8}) holds, but violates the strong one, while the one on the 
8-brane $\Phi_{2} = 0$ violates all the three energy conditions. When $a < - 1$ and $b < - 1$, 
the matter fields on the two 8-branes  all violate the  three energy conditions. However, in
all these four cases, the spacetime singular behavior is similar to the corresponding
5-dimensional cases in the Einstein frame. In particular, in the first three cases the spacetime 
in the four regions and on the 8-branes are always singular, and the corresponding
Penrose diagrams are given, respectively, by Figs. \ref{fig1b}, \ref{fig2b}, \ref{fig3b}, \ref{fig4b}, 
and \ref{fig5b}, but now each point in these figures now represents a 8-dimensional spatial
space. In the last case, in which the matter fields on the two 8-branes violate all the energy
conditions, the spacetime is free of any kind of spacetime singularities, either in Regions $I - 
IV$ or on the two 8-branes, and the corresponding Penrose diagram is given  by Fig. \ref{fig6b}.
Therefore, all the above results seemingly indicate that violating the energy conditions is a 
necessary condition for spacetimes of colliding branes to be non-singular.

\section{Conclusions}
\renewcommand{\theequation}{6.\arabic{equation}}
\setcounter{equation}{0}

In this paper, we have first developed the general formulas to describe the collision of two
timelike (D-1)-branes without $Z_{2}$ symmetry in a D-dimensional effective theory, obtained 
from the toroidal compactification of the Neveu-Schwarz/Neveu-Schwarz (NS-NS) sector  in 
(D+d) dimensions. 
Applying the formulas to the case  $D = 5 = d$ for a class of spacetimes, In Section III
we have obtained explicitly the field equations both outside and on the 3-branes in terms 
of distributions. In Section IV, we have considered a class of exact solutions that represents
the collision of two 3-branes in the Einstein frame, and  studied their local and global 
properties in details. We have found, among other things, that the collision in general 
ends up with the formation of spacetime singularities, due to the mutual focus of the 
colliding branes, although non-singular spacetime also exist, with the price that both of 
the two branes violate all the energy conditions, weak, strong and dominant. Similar
conclusions hold also in the 5-dimensional string frame. This has been done in Section V. 
In Section VI, after lifted the solutions to 10-dimensional spacetimes, we have found that
the corresponding solutions represent the collision of two timelike 8-branes without
$Z_{2}$ symmetry. In some cases the two 8-branes satisfy the weak and dominant energy
conditions, while in other case, they do not. But, in all these cases the strong energy
condition is always violated. The formation of spacetime singularities due to the mutual
focus of the two colliding branes occurs in general, although the non-singular cases
also exist with the price that both of the two branes violate all the three energy
conditions. The spacetime singular behavior is similar in the 5-dimensional effective
theory to that of 10-dimensional string theory.  

In this paper, we have ignored the dilaton $\hat{\Phi}$ and the three-form field
$\hat{H}_{ABC}$. It would be very interesting to see how these fields affect the
formation of the spacetime singularities. In addition, it would also be very interesting
to see what might happen if the branes are allowed to collide more than one time.

\section*{Acknowledgement}
ZCW is supported by the NSFC grant, No. 10703005 and No. 10775119.

\section*{Appendix: Gravitational field equations in the 10-dimensional bulk and
on the 8-branes}
\renewcommand{\theequation}{A.\arabic{equation}}
\setcounter{equation}{0}

For the metric,
\bqn
\lb{a.0}
d^{2}\hat{s}_{10}  &=&  e^{2\hat{\sigma}(t,y)}\left(dt^{2} - dy^{2}\right)
- e^{2\hat{\omega}(t,y)}d\Sigma^{2}_{0} \nb\\
& & - \hat{\phi}^{2}\left(t, y\right) d\Sigma_{z}^{2},
\eqn 
where 
\bq
\lb{a.0a}
d\Sigma^{2}_{0} \equiv \sum^{4}_{p =2}{\left(dx^{p}\right)^{2}},\;\;\;\;
d\Sigma^{2}_{z} \equiv \sum^{5}_{i =1}{\left(dz^{i}\right)^{2}},
\eq
the non-vanishing components of the Einstein tensor are
given by,
\bqn
\lb{a.1}
G^{(10)}_{tt} &=& 3\hat{\omega}_{,t}\left(\hat{\sigma}_{,t} + \hat{\omega}_{,t}\right)
                  + 5 \hat{\psi}_{,t}\left(\hat{\sigma}_{,t} + 3\hat{\omega}_{,t}
		  + 2\hat{\psi}_{,t}\right)\nb\\
               & & -3 \hat{\omega}_{,yy}  - 5 \hat{\psi}_{,yy}
	         - 15\hat{\psi}_{,y}\left(\hat{\omega}_{,y} + \hat{\psi}_{,y}\right)\nb\\
	       & &
		 + \hat{\sigma}_{,y}\left(3\hat{\omega}_{,y} + 5\hat{\psi}_{,y}\right)
		 - 6 {\hat{\omega}_{,y}}^{2},\nb\\
G^{(10)}_{ty} &=& -3\hat{\omega}_{,ty} - 5\hat{\psi}_{,ty}\nb\\
              & & + 3\left(\hat{\sigma}_{,t}\hat{\omega}_{,y}  
	          + \hat{\sigma}_{,y}\hat{\omega}_{,t}
		  - \hat{\omega}_{,t}\hat{\omega}_{,y}\right)\nb\\
	      & & + 5\left(\hat{\sigma}_{,t}\hat{\psi}_{,y}  
	          + \hat{\sigma}_{,y}\hat{\psi}_{,t}
		  - \hat{\psi}_{,t}\hat{\psi}_{,y}\right),\nb\\
G^{(10)}_{yy} &=& -3 \hat{\omega}_{,tt}  - 5 \hat{\psi}_{,tt}
	         - 15\hat{\psi}_{,t}\left(\hat{\omega}_{,t} + \hat{\psi}_{,t}\right)\nb\\
	       & &
		 + \hat{\sigma}_{,t}\left(3\hat{\omega}_{,t} + 5\hat{\psi}_{,t}\right)
		 - 6 {\hat{\omega}_{,t}}^{2}\nb\\
	       & & + 3\hat{\omega}_{,y}\left(\hat{\sigma}_{,y} + \hat{\omega}_{,y}\right)\nb\\
	       & &
                  + 5 \hat{\psi}_{,y}\left(\hat{\sigma}_{,y} + 3\hat{\omega}_{,y}
		  + 2\hat{\psi}_{,y}\right),\nb\\
G^{(10)}_{pq} &=& \delta_{pq}e^{2\left(\hat{\omega} - \hat{\sigma}\right)}
                  \left[\hat{\sigma}_{,yy} + 2\hat{\omega}_{,yy}  
                  + 5 \hat{\psi}_{,yy}\right.\nb\\
	      & & + 5 \hat{\psi}_{,y}\left(2\hat{\omega}_{,y} + 3\hat{\psi}_{,y}\right)
		  + 3 {\hat{\omega}_{,y}}^{2}\nb\\
	      & &- \left(\hat{\sigma}_{,tt}  +2  \hat{\omega}_{,tt}  + 5 \hat{\psi}_{,tt}\right.\nb\\
	      & & \left. + 3 {\hat{\omega}_{,t}}^{2}
	          + 5 \hat{\psi}_{,t}\left(2\hat{\omega}_{,t} + 3\hat{\psi}_{,t}\right)\right],\nb\\
G^{(10)}_{ij} &=& \delta_{ij}e^{2\left(\hat{\psi} - \hat{\sigma}\right)}\left[\hat{\sigma}_{,yy} 
                  + 3\hat{\omega}_{,yy}  
                  + 4 \hat{\psi}_{,yy}\right.\nb\\
	      & & +  2\hat{\psi}_{,y}\left(6\hat{\omega}_{,y} + 5\hat{\psi}_{,y}\right)
		  + 6{\hat{\omega}_{,y}}^{2}\nb\\
	      & &- \left(\hat{\sigma}_{,tt}  +3  \hat{\omega}_{,tt}  + 4 \hat{\psi}_{,tt}\right.\nb\\
	      & & \left. - 10 {\hat{\psi}_{,t}}^{2}
	          + 6 \hat{\omega}_{,t}\left(\hat{\omega}_{,t} + 2\hat{\psi}_{,t}\right)\right], 	 
\eqn
where $p, \; q = 2,\; 3,\;4$ and $i, \; j = 1, ..., \; 5$, and $\hat{\psi} \equiv
\ln\left(\hat{\phi}\right)$.

\subsection{Field Equations on the hypersurface $\Phi_{1} = 0$}

Following Section III.B.1, it can be shown that the derivatives of  any given function $F(t, y)$, 
which is $C^{0}$ across the hypersurface $\Phi_{1} = 0$ and at least $C^{2}$ in the regions $\Phi_{1} 
> 0 $ and $\Phi_{1} > 0 $, are given by Eq.(\ref{3.24}) but now with $N$ being replaced by $\hat{N}$,
and $n_{a}$ and $u_{a}$ by, respectively, $\hat{n}_{a}$ and $\hat{u}_{a}$, where
\bqn
\lb{a.2}
\hat{n}_{a} &=& \hat{N} \left(\delta^{t}_{a} - a \delta^{y}_{a}\right),\nb\\
\hat{u}_{a} &=& \hat{N} \left(a\delta^{t}_{a} -  \delta^{y}_{a}\right),\nb\\
\hat{N} &\equiv& \frac{e^{\hat{\sigma}^{(1)}}}{\left(a^{2} - 1\right)^{1/2}}. 
\eqn
Hence,  Eq.(\ref{a.1})  can be cast in the form,
\bqn
\lb{a.3}
G^{(10)}_{ab} &=& G^{(10) +}_{\;ab} H\left(\Phi_{1}\right) 
+  G^{(10)-}_{\; ab} \left[1 - H\left(\Phi_{1}\right)\right]\nb\\
& & + G^{(10)Im}_{\;ab}\delta\left(\Phi_{1}\right),
\eqn 	 
where $G^{(10) +}_{\;ab}\; \left(G^{(10)-}_{\; ab}\right)$ is the Einstein tensor calculated in
the region $\Phi_{1} > 0\; \left(\Phi_{1} < 0\right)$, and $G^{(10)Im}_{\;ab}$ denotes
the distribution of the Einstein tensor on the hypersurface $\Phi_{1} = 0$, which has the following
non-vanishing components,
\bqn
\lb{a.4}
G^{(10) Im}_{\;tt} &=& a^{2} \hat{N} \left(3\left[\hat{\omega}_{n}\right]^{-} 
             + 5\left[\hat{\psi}_{n}\right]^{-}\right),\nb\\
G^{(10) Im}_{\;ty} &=& - a  \hat{N} \left(3\left[\hat{\omega}_{n}\right]^{-} 
             + 5\left[\hat{\psi}_{n}\right]^{-}\right),\nb\\
G^{(10) Im}_{\;yy} &=&  \hat{N} \left(3\left[\hat{\omega}_{n}\right]^{-} 
             + 5\left[\hat{\psi}_{n}\right]^{-}\right),\nb\\
G^{(10) Im}_{pq} &=& -  \delta_{pq} \hat{N}^{-1} e^{2\hat{\omega}^{(1)}}
                     \left(\left[\hat{\sigma}_{n}\right]^{-} \right.\nb\\
		 & & \;\;\;\;\;\;\left. + 2\left[\hat{\omega}_{n}\right]^{-}
             + 5\left[\hat{\psi}_{n}\right]^{-}\right),\nb\\	     
G^{(10) Im}_{ij} &=& - \delta_{ij}\hat{N}^{-1}  e^{2\hat{\psi}^{(1)}}
                     \left(\left[\hat{\sigma}_{n}\right]^{-} \right.\nb\\
		 & & \;\;\;\;\;\;\left. + 3\left[\hat{\omega}_{n}\right]^{-}
             + 4\left[\hat{\psi}_{n}\right]^{-}\right).	     
\eqn
Introducing the unit vectors,
\bq
\lb{a.5}
X^{(p)}_{a}  = e^{\hat{\omega}^{(1)}}\delta^{p}_{a},\;\;\;
Z^{(i)}_{a}  = e^{\hat{\psi}^{(1)}}\delta^{i}_{a},
\eq
we find that Eq.(\ref{a.4}) can be cast in the form,
\bqn
\lb{a.6}
G^{(10) Im}_{ab} &=& \kappa^{2}_{10}\left(\hat{\rho}_{u} \hat{u}_{a}\hat{u}_{b}
+ \hat{p}^{X}_{u}\sum^{4}_{p=2}{X^{(p)}_{a}X^{(p)}_{b}}\right.\nb\\
& & \left. +  \hat{p}^{Z}_{u}\sum^{5}_{i=1}{Z^{(i)}_{a}Z^{(i)}_{b}}\right),
\eqn
where
\bqn
\lb{a.7}
\hat{\rho}_{u} &=&\frac{1}{\hat{N}\kappa^{2}_{10}} \left(3\left[\hat{\omega}_{n}\right]^{-} 
             + 5\left[\hat{\psi}_{n}\right]^{-}\right),\nb\\
\hat{p}^{X}_{u} &=& \frac{1}{\hat{N}\kappa^{2}_{10}}\left(\left[\hat{\sigma}_{n}\right]^{-}  
                    + 2\left[\hat{\omega}_{n}\right]^{-}
                    + 5\left[\hat{\psi}_{n}\right]^{-}\right),\nb\\
\hat{p}^{Z}_{u} &=& \frac{1}{\hat{N}\kappa^{2}_{10}} \left(\left[\hat{\sigma}_{n}\right]^{-}  
                    + 3\left[\hat{\omega}_{n}\right]^{-}
                    + 4\left[\hat{\psi}_{n}\right]^{-}\right).
\eqn
 
\subsection{Field Equations on the hypersurface $\Phi_{2} = 0$}

Similarly, it can be shown that,  crossing the hypersurface $\Phi_{2} = 0$, 
Eq.(\ref{a.1})  can be cast in the form,
\bqn
\lb{a.3a}
G^{(10)}_{ab} &=& G^{(10) +}_{\;ab} H\left(\Phi_{2}\right) 
+  G^{(10)-}_{\; ab} \left[1 - H\left(\Phi_{2}\right)\right]\nb\\
& & + G^{(10)Im}_{\;ab}\delta\left(\Phi_{2}\right),
\eqn 	 
but now $G^{(10) +}_{\;ab}\; \left(G^{(10)-}_{\; ab}\right)$ is the Einstein tensor calculated in
the region $\Phi_{2} > 0\; \left(\Phi_{2} < 0\right)$, and $G^{(10)Im}_{\;ab}$ denotes
the distribution of the Einstein tensor on the hypersurface $\Phi_{2} = 0$, which can be written in the form,
\bqn
\lb{a.6a}
G^{(10) Im}_{ab} &=& \kappa^{2}_{10}\left(\hat{\rho}_{v} \hat{v}_{a}\hat{v}_{b}
+ \hat{p}^{X}_{v}\sum^{4}_{p=2}{X^{(p)}_{a}X^{(p)}_{b}}\right.\nb\\
& &\left. +  \hat{p}^{Z}_{v}\sum^{5}_{i=1}{Z^{(i)}_{a}Z^{(i)}_{b}}\right),
\eqn
where
\bqn
\lb{a.7a}
\hat{\rho}_{v} &=&\frac{1}{\hat{L}\kappa^{2}_{10}} \left(3\left[\hat{\omega}_{l}\right]^{-} 
             + 5\left[\hat{\psi}_{l}\right]^{-}\right),\nb\\
\hat{p}^{X}_{v} &=& \frac{1}{\hat{L}\kappa^{2}_{10}}\left(\left[\hat{\sigma}_{l}\right]^{-}  
                    + 2\left[\hat{\omega}_{l}\right]^{-}
                    + 5\left[\hat{\psi}_{l}\right]^{-}\right),\nb\\
\hat{p}^{Z}_{v} &=& \frac{1}{\hat{L}\kappa^{2}_{10}} \left(\left[\hat{\sigma}_{l}\right]^{-}  
                    + 3\left[\hat{\omega}_{l}\right]^{-}
                    + 4\left[\hat{\psi}_{l}\right]^{-}\right),
\eqn
and
\bqn
\lb{a.5a}
X^{(p)}_{a}  &=& e^{\hat{\omega}^{(2)}}\delta^{p}_{a},\;\;\;\;\;\;\;\;\;\;
Z^{(i)}_{a}  = e^{\hat{\psi}^{(2)}}\delta^{i}_{a},\nb\\
\hat{l}_{a} &=& \hat{L} \left(\delta^{t}_{a} + b \delta^{y}_{a}\right),\;\;\;
\hat{v}_{a} = \hat{L} \left(b\delta^{t}_{a} +  \delta^{y}_{a}\right),\nb\\
\hat{L} &\equiv& \frac{e^{\hat{\sigma}^{(2)}}}{\left(b^{2} - 1\right)^{1/2}}. 
\eqn

\end{document}